\documentclass[prd,showpacs,eqsecnum]{revtex4}

\usepackage[centertags]{amsmath}
\usepackage{amssymb}
\usepackage{latexsym}
\usepackage{enumerate}
\usepackage{graphicx}
\usepackage{mathrsfs}
\usepackage{hyperref}
\usepackage{stmaryrd}

\newcommand{\bi}{\begin{itemize}}
\newcommand{\ei}{\end{itemize}}

\newcommand{\be}{\begin{equation}}
\newcommand{\ee}{\end{equation}}

\renewcommand{\l}{\left(}
\renewcommand{\r}{\right)}
\renewcommand{\a}{\alpha}
\renewcommand{\b}{\beta}
\newcommand{\g}{\gamma}

\renewcommand{\d}{\delta}

\newcommand{\La}{\Lambda}
\newcommand{\la}{\lambda}
\renewcommand{\O}{\Omega}
\renewcommand{\o}{\omega}
\renewcommand{\th}{\theta}

\newcommand{\q}{\quad}
\newcommand{\s}{\sigma}
\newcommand{\vp}{\varphi}

\newcommand{\pa}{\partial}

\begin{document}

\title{Gravitational perturbations and metric reconstruction: 
Method of extended homogeneous solutions applied to 
eccentric orbits 
on a Schwarzschild black hole} 

\author{Seth Hopper}
\email{hoppese@physics.unc.edu}
\affiliation{Department of Physics and Astronomy, University of North 
Carolina, Chapel Hill, North Carolina 27599}
\author{Charles R. Evans}
\email{evans@physics.unc.edu}
\affiliation{Department of Physics and Astronomy, University of North 
Carolina, Chapel Hill, North Carolina 27599}

\begin{abstract}
We calculate the gravitational perturbations produced by a small mass in 
eccentric orbit about a much more massive Schwarzschild black hole and 
use the numerically computed perturbations to solve for the 
metric.  The calculations are initially made in the frequency domain and 
provide Fourier-harmonic modes for the gauge-invariant master functions 
that satisfy inhomogeneous versions of the Regge-Wheeler and Zerilli 
equations.  These gravitational master equations have specific singular 
sources containing both delta function and derivative-of-delta function 
terms.  We demonstrate in this paper successful application of the method 
of extended homogeneous solutions, developed recently by Barack, Ori, and 
Sago, to handle source terms of this type.  The method allows transformation 
back to the time domain, with exponential convergence of the partial mode 
sums that represent the field.  
This rapid convergence holds even in the region of $r$ traversed by the 
point mass and includes the time-dependent location of the point mass 
itself.  We present numerical results of mode calculations for certain 
orbital parameters, including highly accurate energy and angular momentum 
fluxes at infinity and at the black hole event horizon.  We then address 
the issue of reconstructing the metric perturbation amplitudes from the 
master functions, the latter being weak solutions of a particular form to 
the wave equations.  The spherical harmonic amplitudes that represent the 
metric in Regge-Wheeler gauge can themselves be viewed as weak solutions. 
They are in general a combination of (1) two differentiable solutions that 
adjoin at the instantaneous location of the point mass (a result that has 
order of continuity $C^{-1}$ typically) and (2) (in some cases) a delta 
function distribution term with a computable time-dependent amplitude.

\end{abstract}

\pacs{04.25.dg, 04.30.-w, 04.25.Nx, 04.30.Db}

\maketitle

\section{Introduction}

\label{intro}

Considerable research on the two-body problem in general relativity has 
been fostered over the past decade by the prospects of detecting 
gravitational radiation from extreme-mass-ratio binaries.  The general 
relativistic two-body problem is notoriously difficult, as it involves 
dynamics of the motion of the bodies and of the gravitational field itself.  
Gravitational wave emission carries away energy and angular momentum from 
the orbit, leading to inspiral and eventual merger.  The future joint 
NASA-ESA LISA mission \cite{ESA} is expected to detect between
tens and thousands of 
such extreme-mass-ratio inspirals (EMRIs)--binaries composed of a compact 
object ($\mu \sim 1-50 M_{\odot}$) in orbit about a supermassive Kerr black 
hole ($M \sim 10^5 - 10^7 M_{\odot}$) out to cosmological distances ($z\sim 1$) 
\cite{Barack_2009}.  The small mass ratio 
$10^{-7} \lesssim \mu/M \lesssim 10^{-3}$ of expected astrophysical 
sources \cite{NASA} implies a gradual change in orbital parameters, with 
$\gtrsim 10^5$ wave periods as the binary evolves through the LISA 
passband ($10^{-4} - 10^{-2}$ Hz).  Detailed theoretical calculations will 
aid in both detection of EMRI gravitational wave signals and in 
determination of the source's physical parameters.

Quite apart from the prospects of astrophysical observation, this problem 
is one of intrinsic interest in theoretical physics.  Of the various 
possibilities, the physically simplest compact binary is one composed of 
two black holes.  Such a system eliminates the complications of stellar 
microphysics and reduces the problem to a minimum parameter set.  In 
approaching the problem mathematically, the extreme mass-ratio and gradual 
orbital evolution is of benefit theoretically, allowing black hole 
perturbation theory to be used.  Furthermore, the 
small mass ratio allows even the black hole structure of the small mass to 
be ignored (at lowest order), restoring a point-like (particle) behavior 
\cite{Poisson_2004} on length scales that are large compared to $\mu$ 
and thereby simplifying the perturbation problem. 

The perturbation problem proceeds in stages.  At the outset the motion of
the particle is taken as a geodesic ($\mu / M \rightarrow 0$, or zeroth order) 
on the background spacetime.  The first-order (in $\mu / M$) gravitational 
field perturbation is then computed, yielding a new metric 
${\rm g}_{\mu\nu} = g_{\mu\nu} + p_{\mu\nu}$ that corrects the background
metric $g_{\mu\nu}$.  The gravitational waves in the perturbation $p_{\mu\nu}$ 
carry energy and angular momentum to infinity and down the black hole event
horizon, giving rise to a back reaction or local self-force (SF) on the 
particle that has both conservative and dissipative terms.  Formally, the 
SF depends on gradients of $p_{\mu\nu}$ and acts locally on the particle 
to accelerate it off its background geodesic.  Once the first-order 
correction to the motion is successfully computed, the calculation may 
proceed to second order in the field perturbation (see Pound 
\cite{Pound_2009} for a recent background discussion and an alternative 
formulation).

Yet having idealized the small body as a point particle, the metric
perturbation and SF are found to diverge at the location of 
the particle, and the formal perturbation to the equation of motion is 
meaningless without careful regularization.  This problem is similar 
to the classic SF problem of an accelerating, radiating charge 
in electromagnetic theory in flat spacetime \cite{Dirac_1938}.  Two 
pivotal papers, by Mino, Sasaki, and Tanaka \cite{MiSaTa} and Quinn and 
Wald \cite{QuWa}, showed how the metric perturbation may be separated into 
a divergent, direct part $p_{\mu\nu}^{\rm dir}$ and a finite tail term 
$p_{\mu\nu}^{\rm tail}$, with the latter providing the regularized field 
that makes the SF finite.  
As an alternative, Detweiler and Whiting \cite{DW_2003} proposed decomposing
the metric perturbation into regular $p_{\mu \nu}^{R}$ and singular
$p_{\mu \nu}^{S}$ parts.  Under this interpretation, $p_{\mu \nu}^{R}$
is a solution to the vacuum field equations, but gives rise to the
same SF as $p_{\mu\nu}^{\rm tail}$.

Since then, SF calculations have been made in 
certain special cases \cite{BL_2002, BS_2007, Det_2008, BS_2009, BS_2010}.  
See the review by Barack \cite{Barack_2009}.  Ultimately, the theory aims 
to provide self-consistent SF calculations of arbitrary orbits about Kerr 
black holes.  In this paper, we concern ourselves with a more modest 
goal: demonstrating a complete computation of the radiative gravitational 
perturbations produced by a mass in eccentric orbit on a Schwarzschild 
black hole and reconstruction of the corresponding parts of the perturbed 
metric in Regge-Wheeler gauge.  While we leave for another occasion 
computation of both the nonradiative perturbations and the SF, the accurate 
reconstruction of the radiative parts of the metric, at all locations up 
to and including the point mass, should serve as a starting point for a 
further gauge transformation or alternative regularization technique.  

We note in passing that most work to date computing EMRI evolution has not 
made use of local SF calculation.  Sufficiently adiabatic changes in an 
orbit on Schwarzschild spacetime allow a \emph{balance calculation} 
approach \cite{CKP_1994}, where orbital energy and angular momentum are 
``evolved'' (acausally) to match corresponding gravitational wave fluxes 
through bounding surfaces at large radius and near the horizon.  Much effort 
is ongoing to extend the reach of adiabatic calculations 
\cite{Glampedakis_2002,Hughes_Drasco_2005, Mino_2003}.  Unfortunately, the 
approach only approximates dissipative SF terms and cannot account for 
conservative SF effects.  In any event, the more self-consistent SF approach 
should serve to confirm the validity of these or other approximations.

Perturbation theory for Schwarzschild black holes has a traditional 
formalism pioneered by Regge and Wheeler \cite{RW_1957}, Zerilli 
\cite{Zerilli_1970}, and Vishveshwara \cite{Vish_1970} that uses spherical 
harmonics and the Regge-Wheeler gauge to simplify algebraically the form 
of the metric perturbation.  At each spherical harmonic order there are 
just two \emph{master} functions, $\Psi_{\ell m}^{\rm even}(t,r)$ and 
$\Psi_{\ell m}^{\rm odd}(t,r)$, one for each parity or gravitational 
degree of freedom, which satisfy linear inhomogeneous wave equations in 
$t$ and $r$.  The formalism was improved by Moncrief \cite{Moncrief_1974} 
and colleagues \cite{CPM_1978, CPM_1979}, making use instead of 
gauge-invariant master functions that satisfy similar wave equations.
Recently Martel and Poisson \cite{MP_2005} have placed the theory in 
both a gauge-invariant and covariant form. 

For perturbations of Kerr black holes, Teukolsky \cite{Teukolsky_1973} 
developed a formalism based on Newman-Penrose curvature scalars and 
spin-weighted spheroidal harmonics.  In the frequency domain the radial 
part is a single (complex) master equation \cite{Chandra_1983}, which 
can, of course, be applied to a Schwarzschild hole as well \cite{CKP_1994, 
Sasaki_2003}. 

An alternative to the Regge-Wheeler-Zerilli (RWZ) approach has recently been 
advanced by Barack and Lousto \cite{BL_2005}.  They propose directly 
evolving the ten spherical harmonic amplitudes that describe the metric 
perturbation in Lorenz (or harmonic) gauge.  In this direct metric perturbation 
approach, the equations separate into even- and odd-parity sectors, yet 
still involve systems of seven and three coupled equations, respectively.  
Barack and Sago \cite{BS_2007, BS_2010} have used the formalism to compute 
the time evolution of metric perturbations generated by circular and 
eccentric orbits on Schwarzschild, along with the resulting SF 
components.

The RWZ and direct metric perturbation approaches each 
have advantages and disadvantages.  The direct metric perturbation formalism 
yields directly what one wants as an input to a SF calculation, namely 
the metric itself in Lorenz gauge.  In a time domain calculation,
as so far employed, it has the disadvantage of requiring 
simultaneous solution of a large set of coupled partial differential 
equations (PDE's).  Anticipating the 
subtraction involved in the SF regularization, Barack, Lousto, 
and Sago have built a fourth-order convergent finite difference code to 
compute the modes to sufficient accuracy.  In contrast, the RWZ 
approach has the advantage that only a single uncoupled wave 
equation need be solved for each mode and parity.  Unfortunately, an 
added step is required to reconstruct the metric from the mode solutions.  
Moreover, the reconstruction involves terms that are singular at the 
particle location and the simplest reconstruction yields the metric 
perturbation in Regge-Wheeler gauge \cite{Martel_2004,Lousto_2005}.  
Finally, the RWZ approach provides only the radiative ($\ell \geq 2$) 
parts of the perturbation and the nonradiative modes ($\ell = 0,1$) must
be derived by separate means.

In this paper we opt for using the gauge-invariant RWZ approach detailed 
by Martel and Poisson \cite{MP_2005}, and adopt the Zerilli-Moncrief 
$\Psi^{\rm ZM}_{\ell m} = \Psi^{\rm even}_{\ell m}$ and 
Cunningham-Price-Moncrief $\Psi^{\rm CPM}_{\ell m} = \Psi^{\rm odd}_{\ell m}$ 
master functions for even and odd-parity, respectively.  Our use of this 
relatively standard method is augmented, though, by a new technique 
that enables accurate reconstruction of the corresponding parts of the 
metric in Regge-Wheeler gauge.  We leave for a later occasion our own 
consideration of the monopole and dipole terms (which are essential to a SF 
calculation) and instead direct attention to discussion by Detweiler and 
Poisson \cite{DP_2003} and recent successful numerical implementation 
by Barack and Sago \cite{BS_2010}.

The master functions can be obtained directly by numerical evolution 
(solution of PDE's) in the time domain (TD) (see e.g., \cite{BB_2000,BL_2002,
Martel_2004,Haas_2007,BS_2007,SL_2008,BS_2009, BS_2010}) or by numerical 
integration of ordinary differential equations (ODE's) for the Fourier 
modes in the frequency domain (FD) (see e.g., \cite{CKP_1994,Burko_2000,
DMW_2003,BOS}).  Each method has strengths and weaknesses.  
TD calculations require solving just one equation for each 
$\ell, m$ mode and time dependence of the subsequently reconstructed 
metric and SF is of direct interest.  Disadvantages of TD calculations 
include (1) modeling the discontinuous source movement through the finite 
difference grid \cite{BL_2005,BS_2010}; (2) numerical stability of PDE 
evolution; (3) difficulty devising numerical schemes of adequately small 
truncation error; and (4) challenges in posing outgoing wave boundary 
conditions at finite radius.  In contrast, in FD 
calculations (1) the numerical errors tend to be much smaller (i.e., by 
solving ODE's); (2) outgoing wave boundary conditions are handled 
mode-by-mode and extrapolated to infinity and to the black hole event 
horizon; and (3) the discontinuous source presents few difficulties in
computing (at least) the Fourier mode functions $R_{\ell m n}(r)$.  
However, FD methods require, for eccentric orbits, computing and summing 
over numerous harmonics $n$ of the radial libration frequency $\Omega_r$ 
for each $\ell, m$ and transformation to the TD is nontrivial given the 
singular source terms.

Barack, Ori, and Sago (BOS) \cite{BOS} highlighted the latter difficulty.  
They used the model problem of a scalar field $\Phi (t,r,\theta,\varphi)$ 
generated by a scalar point charge in eccentric orbit on Schwarzschild.  
The spherical harmonic modes $\phi_{\ell m}(t,r) = r \Phi_{\ell m}(t,r)$ 
satisfy a wave equation with a singular source, 
$S^{\rm scalar}_{\ell m}(t,r) = C_{\ell m}(t,r) \, \d [r-r_p(t)]$.  Here 
$C_{\ell m}(t,r)$ is some smooth function and $r = r_p(t)$ describes the 
radial libration of the particle's worldline between two turning points.  
In the FD, ODE's are solved for the Fourier-harmonic modes $R_{\ell mn}(r)$.  
These mode functions are, at each point $r$, Fourier series coefficients.  
The resulting Fourier series converges for the piecewise continuous 
($C^0$) $\phi_{\ell m}(t,r)$ but the singular nature of the source $S$ 
makes $\phi_{\ell m}(t,r)$ converge slowly in the region traversed by the 
point charge.  The radial derivative $\pa_{r} \phi_{\ell m}$ is however 
discontinuous at $r=r_p(t)$ and its Fourier series only converges, in the 
usual sense~\cite{champeney_1989}, almost everywhere.  The attempt to 
assemble the radial derivative from the Fourier series is plagued by the 
Gibbs phenomenon; the series converges to the mean value at the discontinuity 
and the series ``overshoots'' and fails to converge properly in the limit 
as both $n\rightarrow \infty$ and $r\rightarrow r_p(t)^{\pm}$.

BOS circumvented the difficulty with a new \emph{method of extended 
homogeneous solutions}.  In brief, they use FD analysis to find 
Fourier-harmonic mode solutions to the homogeneous equation, valid outside 
and on either side of the source libration region.  The associated Fourier 
series converge exponentially fast to homogeneous solutions of the TD 
wave equation.  They then analytically extend both homogeneous TD solutions 
into the source libration region up to the instantaneous position of the 
point charge.  Summed to adequately high order, the two homogeneous solutions 
match in value at $r_p(t)$, as expected.  With the field represented in 
this way, the left and right derivatives can be accurately determined.  
BOS argued that the method should work for other problems with similar wave 
equations, including the Teukolsky equation.

We show in this paper that the method can indeed be extended to the case 
of gravitational perturbations computed in the RWZ formalism, and apply 
the method to a large set of Fourier-harmonic modes stemming from a mass in 
eccentric orbit on Schwarzschild.  (Note that Barack and Sago \cite{BS_2010} 
previously implemented this method in the gravitational case but only for 
the $\ell = 0,1$ modes in Lorenz gauge.)  An important distinction arises: in 
the gravitational case the source distribution in the Regge-Wheeler gauge
contains both delta function 
and derivative-of-delta function terms,
\be
\label{eq:SMartelForm}
S_{\ell m}(t,r) = G_{\ell m}(t,r) \, \d [r-r_p(t)] + 
F_{\ell m}(t,r) \, \d' [r-r_p(t)] ,
\ee
with $G_{\ell m}(t,r)$ and $F_{\ell m}(t,r)$ being smooth functions.  As a 
consequence the master functions have a jump discontinuity at $r=r_p(t)$ 
(referred to sometimes as a $C^{-1}$ function).  The resulting extension
of the homogeneous solutions, $\Psi_{\ell m}^{+}$ and
$\Psi_{\ell m}^{-}$, written as
\be
\label{eq:weakPsi}
\Psi_{\ell m}(t,r) = \Psi_{\ell m}^{+}(t,r) \, \theta[r-r_p(t)] + 
\Psi_{\ell m}^{-}(t,r) \, \theta[r_p(t)-r] ,
\ee
where $\theta[r-r_p(t)]$ is the Heaviside function, is a type of 
\emph{weak solution} to the inhomogeneous master equation.  Thus in the 
gravitational case in RWZ gauge the difficulty with local convergence occurs 
with the master function itself.  We show that the use of distributions, or 
generalized functions \cite{Lighthill_1958}, makes possible separate 
analytic calculation of the expected jumps in value and slope of 
$\Psi_{\ell m}$.  We further demonstrate that the metric perturbation can 
be accurately numerically computed, including the time dependent magnitudes 
of delta function terms that appear in some of the metric amplitudes in 
Regge-Wheeler gauge.

This paper is organized as follows.  In Sec.~\ref{analytic} we 
briefly outline the general mathematical problem of using FD techniques 
to solve for perturbations in the RWZ formalism.  We also 
review the standard parameterization of eccentric orbits.  Sec.~\ref{EHS} 
concerns the method of extended homogeneous solutions.  We first review 
BOS's solution for the scalar field case.  We show then 
our treatment of more general source terms and extension of the method to 
gravitational perturbations.  Sec.~\ref{modeIntegration} provides numerical 
results on the computed Fourier-harmonic mode functions, including 
convergence tests and calculation of radiated gravitational wave energy 
and angular momentum.  In particular, the energy and angular momentum 
fluxes are shown to agree with past published values.  More importantly, 
the method is shown to provide a solution to the field and its derivatives 
that is convergent exponentially fast everywhere.  Then in 
Sec.~\ref{reconstruct}, we show that the equations which allow the metric 
to be obtained from the master functions, along with an understanding of 
the form of the weak solutions for $\Psi_{\ell m}^{\rm even}$ and 
$\Psi_{\ell m}^{\rm odd}$, can be used to determine both the smooth and 
distributional parts of the metric.  App.~\ref{generalized} discusses 
fully evaluated forms of distributional source terms. 
App.~\ref{GF_Evaluated} gives the details of such source terms
for our case of eccentric orbits on Schwarzschild.  In App.~\ref{MP} we 
concisely summarize the metric perturbation formalism in the Regge-Wheeler 
gauge.  We show the construction of gauge-invariant master functions of 
each parity, and provide the spherical harmonic decomposition of the 
Einstein equations and Bianchi identities.  App.~\ref{asympExp} concludes 
this paper with a brief discussion of asymptotic expansions used to set 
boundary conditions on the mode functions at large $r$.  

Throughout this paper we use the sign conventions and notation of Misner, 
Thorne, and Wheeler \cite{MTW} and use units in which $c = G = 1$.  We 
use Schwarzschild coordinates $x^{\mu} = (t,r,\th ,\varphi )$ except as 
otherwise indicated.

\section{Background on the standard RWZ approach to  
gravitational perturbations in the frequency domain}
\label{analytic}

In this section we briefly summarize both the standard notation for 
parameterizing bound orbits on Schwarzschild and the usual approach to 
computing gravitational perturbations using the Regge-Wheeler-Zerilli (RWZ)
formalism
in the frequency domain (FD).  The description of the geodesic motion on 
the background, in terms of various curve functions, is used throughout the 
rest of the paper.  The standard FD analysis provides the notation for 
describing the Fourier-harmonic modes, and their normalization, and sets 
the stage for discussion in Sec.~\ref{EHS} of how gravitational 
perturbations can be returned successfully to the time domain (TD).  Here,
and throughout this paper, we use a subscript $p$ to indicate evaluation along
the worldline of the particle.

\subsection{Bound orbits on a Schwarzschild black hole}
\label{orbits}

Consider bound timelike geodesic motion around a Schwarzschild black hole 
(i.e., $\mu \rightarrow 0$).  We may for the nonce use proper time $\tau$ 
to parameterize the geodesic, 
$x_p^{\mu}(\tau) 
= \left[ t_p(\tau),r_p(\tau),\th_p(\tau),\varphi_p(\tau) \right]$, with
the associated four-velocity $u^{\mu} = dx_p^{\mu}/d\tau$.  On Schwarzschild
we take $\theta_p(\tau)=\pi/2$ without loss of generality.  The geodesic 
equations yield immediate first integrals and allow the trajectory to be 
described by the conserved energy ${\cal{E}}$ and angular momentum 
${\cal{L}}$ per unit mass.  Alternatively, we can choose the (dimensionless) 
semi-latus rectum $p$ and the eccentricity $e$ as orbital parameters 
(c.f., \cite{CKP_1994,BS_2010}).  A third choice would be use of the 
periapsis and apapsis, $r_{\rm min}$ and $r_{\rm max}$.  We will find all 
of these useful in what follows.  The latter two parameter pairs are 
related to each other by 
\be
p \equiv \frac{2 r_{\rm max} r_{\rm min}}{M (r_{\rm max} + r_{\rm min})} , 
\q \q
e \equiv \frac{r_{\rm max} - r_{\rm min}}{r_{\rm max} + r_{\rm min}},
\ee
or inversely
\be
r_{\rm max} = \frac{pM}{1-e}, 
\q \q
r_{\rm min} = \frac{pM}{1+e}.
\ee
The specific energy and angular momentum are related to $p$ and $e$ 
by \cite{CKP_1994}
\be
{\cal{E}}^2 = \frac{(p-2-2e)(p-2+2e)}{p(p-3-e^2)}, 
\q \q
{\cal{L}}^2 = \frac{p^2 M^2}{p-3-e^2}.
\ee
The geodesic equations provide the following differential equations for 
the orbital motion and for the time dependence of the four-velocity, 
\be
\label{eq:fourVelocity}
\frac{dt_p}{d\tau} = u^t = \frac{{\cal{E}}}{f_{p}},  
\q \q
\frac{d\varphi_p}{d\tau} = u^{\varphi} = \frac{{\cal{L}}}{r_p^2} ,
\q \q
\l \frac{dr_p}{d\tau} \r^2 =\l u^r \r^2 = {\cal E}^2-U^2_{p}, 
\ee
where
\be
f(r) \equiv 1 - \frac{2M}{r},
\q \q
U^2(r,{\cal{L}}^2) \equiv f \l 1 + \frac{{\cal{L}}^2}{r^2} \r .
\ee

For purposes of numerical integration there is another curve parameter, 
originally devised by Darwin \cite{Darwin}, that proves useful.  Here one 
introduces a phase angle $\chi$ that is related to the radial 
position on the orbit by the Keplerian-appearing form
\be
r_p \l \chi \r = \frac{pM}{1+ e \cos \chi} .
\ee
Of course, in the relativistic case $\chi$ differs from the true anomaly 
$\varphi$.  The orbit goes through one radial libration for each change 
$\Delta\chi = 2\pi$.  The use of $\chi$ eliminates 
singularities in the differential equations at the turning points 
\cite{CKP_1994}.  Note 
that at $\chi = 0$, $r_p = r_{\rm min}$ and at 
$\chi = \pi$, $r_p = r_{\rm max}$.  (Also note that in this section we are 
content with making a slight abuse of notation in jumping from $r_p(\tau)$ 
to $r_p(\chi)$, before ultimately settling on $r_p(t)$.)  In terms of $\chi$
the equations are
\be
\label{eq:dtdChi}
\frac{dt_p}{d \chi} = \frac{p^2 M}{(p - 2 - 2 e \cos \chi) (1 + e \cos \chi)^2}
\left[ \frac{(p-2)^2 - 4 e^2}{p - 6 - 2 e \cos \chi} \right]^{1/2},
\ee
\be
\frac{d \varphi_p}{d\chi} 
= \left[\frac{p}{p - 6 - 2 e \cos \chi}\right]^{1/2} ,
\ee
and
\be
\frac{d\tau_p}{d \chi} = \frac{M p^{3/2}}{(1 + e \cos \chi)^2} 
\left[ \frac{p - 3 - e^2}{p - 6 - 2 e \cos \chi} \right]^{1/2} .
\ee
We use Eq.~(\ref{eq:dtdChi}) to derive the fundamental frequency  and
period of radial motion,
\be
\label{eq:O_r}
\O_r \equiv   \frac{2 \pi}{T_r},
\q \q 
T_r \equiv \int_{0}^{2 \pi} \l \frac{dt_p}{d\chi} \r d \chi.
\ee
It is also of importance to have the average rate at which the azimuthal 
angle advances, found by averaging the angular frequency $d \varphi_p / dt$ 
over a radial libration via
\be
\label{eq:O_phi}
\O_\varphi \equiv \frac{1}{T_r} \int_{0}^{T_r} \l \frac{d \varphi_p}{dt} \r dt .
\ee
While $T_r$ represents the lapse of coordinate time in a radial libration, 
the time $T_\varphi = 2 \pi / \O_\varphi$ has no particular physical 
significance \cite{Schmidt_2002}.  
Finally, because wave equation source functions contain 
terms like $\d [r-r_p(t)]$ and $\d' [r-r_p(t)]$, we have need of derivatives
of $r_p(t)$,
\be
\label{eq:removeRDot}
\dot r_p^2(t) = f_{p}^2 - \frac{f_{p}^2}{{\cal{E}}^2} U^2_{p} , 
\q \q
\ddot r_p (t) =  \frac{2M f_{p}}{r_{p}^2} - \frac{f_{p}^2}{{\cal{E}}^2 r_{p}^2} 
\left[3 M -  \frac{{\cal{L}}^2}{r_{p}} 
+ \frac{5 M {\cal{L}}^2}{r_{p}^2} \right],
\ee
where we let a dot signify differentiation with respect to coordinate time.

\subsection{The Regge-Wheeler-Zerilli formalism in the frequency domain}

As discussed in the Introduction, we use the RWZ approach to gravitational 
perturbations and use specifically the even-parity Zerilli-Moncrief 
function $\Psi_{\ell m}^{\rm even}$ \cite{Moncrief_1974} and the odd-parity 
Cunningham-Price-Moncrief function $\Psi_{\ell m}^{\rm odd}$ \cite{CPM_1979}.  
See Martel and Poisson \cite{MP_2005} for recent discussion and references 
therein.  Both of these functions satisfy wave equations of the form
\be
\left[ -\frac{\pa^2}{\pa t^2}  + \frac{\pa^2}{\pa r_*^2} 
- V_{\ell}(r) \right] \Psi_{\ell m}(t,r) = S_{\ell m}(t,r) ,
\label{eq:TDwaveEq}
\ee
where $r_* = r + 2M \ln (r/2M - 1)$ is the usual tortoise coordinate. 
The potential used in Eq.~(\ref{eq:TDwaveEq}) is either the Zerilli  
or Regge-Wheeler potential depending on whether the parity is even or odd, 
respectively.

The source terms also depend upon parity but further depend on which specific 
master functions are chosen.  Martel and Poisson gave the 
covariant form of $S_{\ell m}^{\rm even}$ and $S_{\ell m}^{\rm odd}$ (see 
App.~\ref{MP} for these in Schwarzschild coordinates) that are associated 
with the  Zerilli-Moncrief and Cunningham-Price-Moncrief  functions.  Martel 
\cite{Martel_2004} derived the detailed form of $S_{\ell m}^{\rm even}$ for 
a point mass in eccentric orbit.  Sopuerta and Laguna \cite{SL_2008} 
derived the detailed form of $S_{\ell m}^{\rm odd}$ for eccentric orbits
(see also Field et al. \cite{FHL_2009}).  
We give in App.~\ref{GF_Evaluated} detailed expressions for these 
sources in a form that is useful for both mode integrations and metric 
reconstruction.  

In each case the source term has the following general form
\be
\label{eq:gravitySource}
S_{\ell m}(t,r) = \tilde G_{\ell m}(t) \, \d [r - r_p(t)] 
+ \\ \tilde F_{\ell m}(t) \, \d' [r - r_p(t)] ,
\ee
where $\tilde G_{\ell m}(t)$ and $\tilde F_{\ell m}(t)$ are smooth 
(differentiable) functions.  Note that the source, as written here, differs 
from notation originally used by Martel \cite{Martel_2004} (who retained 
smooth functions of $r$ and $t$, as in Eq.~(\ref{eq:SMartelForm})).  Our 
expression uses the delta function, 
and parts integration, to yield a \emph{fully evaluated form} along the 
worldline of the particle (see App.~\ref{generalized}), making 
$\tilde G_{\ell m}(t)$ and $\tilde F_{\ell m}(t)$ unique
functions of time only.

Eq.~(\ref{eq:TDwaveEq}) can be solved 
directly in the TD--an approach that has received much attention 
lately.  In this paper we are interested instead in extending the reach of 
FD analysis, and the balance of this section provides a 
brief review of the standard FD solution.  We note in passing that a 
hybrid approach is possible--using FD analysis for low $\ell$ and $m$ modes
while using TD calculation for high order modes \cite{BLKKB_2008}.

On Schwarzschild, eccentric orbits are typically not closed and therefore 
the motion is not simply periodic as seen by an asymptotic static observer.  
The radial libration is periodic (but not typically sinusoidal) with 
fundamental frequency $\Omega_r$.  The smooth functions $\tilde G_{\ell m}(t)$ 
and $\tilde F_{\ell m}(t)$, which depend upon the particle's radial and 
angular motion, have terms that are periodic with fundamental frequency 
$\Omega_r$, but also involve a term that is proportional to 
$\exp[-im\varphi_p(t)]$.  This latter term comes from restricting the 
spherical harmonics $Y_{\ell m}^*(\theta, \varphi)$ with 
$\d [\varphi -\varphi_p(t)]$.  The function $\varphi_p(t)$ advances with 
an average rate $\Omega_{\varphi}$, but is modulated (in an eccentric orbit)
by a function $\Delta\varphi(t)$ that is periodic with fundamental frequency
$\Omega_r$.  Hence, the source $S_{\ell m}(t,r)$, and therefore the field
$\Psi_{\ell m}(t,r)$, can be represented by a Fourier series with fundamental 
frequency $\Omega_r$, but multiplied by a phase factor that advances linearly 
with rate $\Omega_{\varphi}$.  These fields \emph{would} appear simply 
periodic to an observer whose frame rotates at rate $\Omega_{\vp}$ 
\cite{CKP_1994}.  To a static observer, a given mode $\ell$ and $m$ will 
have a spectrum of harmonics offset by $m\Omega_{\varphi}$; taken together 
the full field will have a two-fold countably infinite frequency spectrum,
\be
\label{eq:omega_mn}
\o = \o_{mn} \equiv m \O_\varphi + n \O_r, \q \q m,n \in \mathbb{Z}.
\ee

Accordingly, the wave equation (\ref{eq:TDwaveEq}) Fourier transforms into
a set of ODE's,
\be
\left[ \frac{d^2}{d r_*^2} -  V_{\ell}(r) + \o^2_{mn} \right]
R_{\ell mn}(r) = Z_{\ell mn}(r) ,
\label{eq:FDwaveEq}
\ee
where $R_{\ell mn}(r)$ and $Z_{\ell mn}(r)$ are Fourier harmonic amplitudes 
\be
R_{\ell mn}(r) \equiv \frac{1}{T_r} \int_0^{T_r} dt \ \Psi_{\ell m}(t,r) 
\, e^{i \o_{mn} t},
\q \q
Z_{\ell mn}(r) \equiv \frac{1}{T_r} \int_0^{T_r} dt \ S_{\ell m}(t,r) 
\, e^{i \o_{mn} t} .
\label{eq:FDSource}
\ee
The series representations of $\Psi_{\ell m} (t,r)$ and $S_{\ell m}(t,r)$ are
\be
\label{eq:FDPsi}
\Psi_{\ell m} (t,r) = \sum_{n=-\infty}^{\infty} 
R_{\ell mn}(r) \, e^{-i \o_{mn} t},
\q \q
S_{\ell m}(t,r) = \sum_{n = -\infty}^\infty 
Z_{\ell mn}(r) \, e^{-i \o_{mn} t},
\ee
and are subject to the usual provisos of Fourier theory regarding for what 
$r$ Eqs.~(\ref{eq:FDPsi}) converge to the original functions.

In order to find the solution to Eq.~(\ref{eq:FDwaveEq}), 
we start by solving the homogeneous version of that equation, obtaining 
two independent solutions.  Using the terminology of Galt'sov \cite{Galtsov} 
(see also \cite{DFH} for a clear presentation of basis modes), the 
$R_{\ell mn}^- (r)$ solution is computed by setting a unit normalized 
``in'' wave boundary condition of
\be
\hat R_{\ell mn}^- (r_* \to -\infty) = e^{-i  \o_{mn} r_*},
\label{eq:in}
\ee
near the horizon.  Similarly, the $R_{\ell mn}^+ (r)$ solution arises from 
setting a unit normalized ``up'' boundary condition of
\be
\hat R_{\ell mn}^+ (r_* \to +\infty) = e^{i  \o_{mn} r_*},
\label{eq:up}
\ee
at large $r_*$.
Formally, these homogeneous solutions are both valid in the entire range 
$2M < r <\infty$.  The standard method of integrating the Green function 
and source (the method of variation of parameters) gives the solution to 
the inhomogeneous equation (\ref{eq:FDwaveEq}),
\be
\label{eq:FDInhomog}
R_{\ell mn} (r) = c^+_{\ell mn}(r) \hat R^+_{\ell mn} (r) 
+ c^-_{\ell mn}(r) \hat R^-_{\ell mn}(r),
\ee
where
\begin{align}
\label{eq:cPM}
c^+_{\ell mn} (r)  &\equiv \frac{1}{W_{\ell mn}} \int_{r_{\rm min}}^r dr'  
\frac{\hat R^-_{\ell mn} (r') Z_{\ell mn} (r')}{ f(r')}, 
&
c^-_{\ell mn} (r)  &\equiv \frac{1}{W_{\ell mn}} \int_r^{r_{\rm max}} dr' 
\frac{\hat R^+_{\ell mn} (r') Z_{\ell mn} (r')}{ f(r')},
\end{align}
and 
\be
W_{\ell mn} \equiv \hat R^-_{\ell mn}
 \frac{d \hat R^+_{\ell mn}}{dr_*} 
- \hat R^+_{\ell mn} 
\frac{d \hat R^-_{\ell mn}}{dr_*},
\ee 
is the Wronskian.  Outside the source libration region, 
Eq.~(\ref{eq:FDInhomog}) reduces to the normalized homogeneous solutions
that are properly connected through the source region,
\begin{align}
\begin{split}
\label{eq:freqHomogSol}
R_{\ell mn}^+ (r) &= 
C_{\ell mn}^+ \hat R_{\ell mn}^+(r), \q r \ge r_{\rm max} , \\
R_{\ell mn}^- (r) &= 
C_{\ell mn}^- \hat R_{\ell mn}^-(r), \q r \le r_{\rm min} ,
\end{split}
\end{align}
where $C^\pm_{\ell mn}$ are the values of $c^\pm_{\ell m n} (r)$ 
evaluated at the ends of the range of the source,
\be
C_{\ell mn}^+  \equiv c_{\ell mn}^+ \l r_{\rm max} \r,
\q \q
C_{\ell mn}^-  \equiv c_{\ell mn}^- \l r_{\rm min} \r.
\label{eq:CNorm}
\ee

\section{The method of extended homogeneous solutions in the 
gravitational case}
\label{EHS}

\subsection{Brief review of Barack, Ori, and Sago's method of extended 
homogeneous solutions}

As a model problem, Barack, Ori, and Sago (BOS) considered the scalar 
field $\Phi$ produced by a scalar point charge in an eccentric orbit on a 
Schwarzschild background.  The spherical harmonic amplitudes 
$\phi_{\ell m}(t,r) = r \Phi_{\ell m}(t,r)$ of the scalar field satisfy 
RWZ-like equations fully analogous to Eq.~(\ref{eq:TDwaveEq}) but with 
source functions that only depend upon a Dirac delta function,
\be
S^{\rm scalar}_{\ell m} 
= C_{\ell m} (t,r) \, \d [r - r_p(t)] .
\ee 
Here $C_{\ell m}(t,r)$ is a smooth function that is derived from the 
particle's point-like charge density $\rho$. 

With a delta function source the amplitudes $\phi_{\ell m}(t,r)$ are
left piecewise continuous ($C^0$) at the instantaneous particle location 
$r_p(t)$ but lose all differentiability there.  BOS argued that this 
behavior, while surmountable in TD calculations, would cause difficulties 
for Fourier synthesis in FD calculations.  As they convincingly demonstrated 
with their first two figures, while $\phi_{\ell m}(t,r)$ converges 
exponentially fast outside the radial libration region, the Gibbs phenomenon 
is responsible for a very slow convergence of $\phi_{\ell m}(t,r)$ between 
$r_{\rm min}$ and $r_{\rm max}$.  Furthermore, the radial derivative
$\partial_r \phi_{\ell m}$ is discontinuous at $r_p(t)$ and suffers the 
full effects of the Gibbs phenomenon--the Fourier series converges to the 
mean value at the discontinuity and partial sums ($-N\le n \le N$) overshoot 
in the limit as both $N \rightarrow \infty$ and $r \rightarrow r_p(t)^{\pm}$.
This behavior is a serious obstacle to straightforward use of FD calculations
in SF regularization.
 
As a solution to this problem, BOS developed the method of extended 
homogeneous solutions (EHS).  Their method involves using the Fourier-harmonic 
modes of the homogeneous equation in the FD to synthesize homogeneous 
solutions $\phi_{\ell m}^- (t,r)$ and $\phi_{\ell m}^+ (t,r)$ to the TD
wave equation.  The Fourier convergence of these homogeneous solutions is 
exponentially rapid.  While these solutions exist in the entire radial 
domain ($2M < r < \infty$), ordinarily $\phi_{\ell m}^- (t,r)$ and 
$\phi_{\ell m}^+ (t,r)$ would be viewed as meaningful in their respective 
source-free regions, $r < r_{\rm min}$ and $r > r_{\rm max}$.  The heart 
of the BOS method lies in extending both of these solutions into the 
region of radial libration up to the instantaneous position of the particle.

BOS demonstrated the method numerically using the monopole term of 
$\Phi$.  A key condition for success of the method is that, as 
$N\rightarrow \infty$ in the partial sums, one finds 
\be
\lim_{r\rightarrow r_p(t)} \phi_{\ell m}^- (t,r) = 
\lim_{r\rightarrow r_p(t)} \phi_{\ell m}^+ (t,r) ,
\ee
as expected analytically.  This was observed numerically and the method as a 
whole converges rapidly since the FD solution of the inhomogeneous 
equation is never summed.  BOS went on to argue that the method could be 
extended to any $\ell$ and $m$ for scalar, electromagnetic, or gravitational 
fields.

\subsection{Application to gravitational perturbations}
\label{jumpCond}

In this section we detail our application of the method to the gravitational 
case in RWZ gauge.  It is worth first observing the magnitude of the
problem to be circumvented.  Given the gravitational source 
(\ref{eq:gravitySource}), and the solution to Eq.~(\ref{eq:FDwaveEq}) 
afforded by Eq.~(\ref{eq:FDInhomog}), the standard approach would represent 
the inhomogeneous solution to the master equation (\ref{eq:TDwaveEq}) by
\be
\label{eq:PsiStd}
\Psi_{\ell m}(t,r) \sim \Psi^{\rm std}_{\ell m}(t,r)
= \sum_{n=-N}^{+N} R_{\ell mn}(r) \, e^{-i \o_{mn} t}, \q \q N \to \infty,
\ee
where we use the $\sim$ to indicate that the equality between the 
actual solution $\Psi_{\ell m}$ and $\Psi_{\ell m}^{\rm std}$ 
holds \emph{almost everywhere} for $N\rightarrow \infty$.

\begin{figure}[h!]
{
	\begin{center}
	{
		\includegraphics{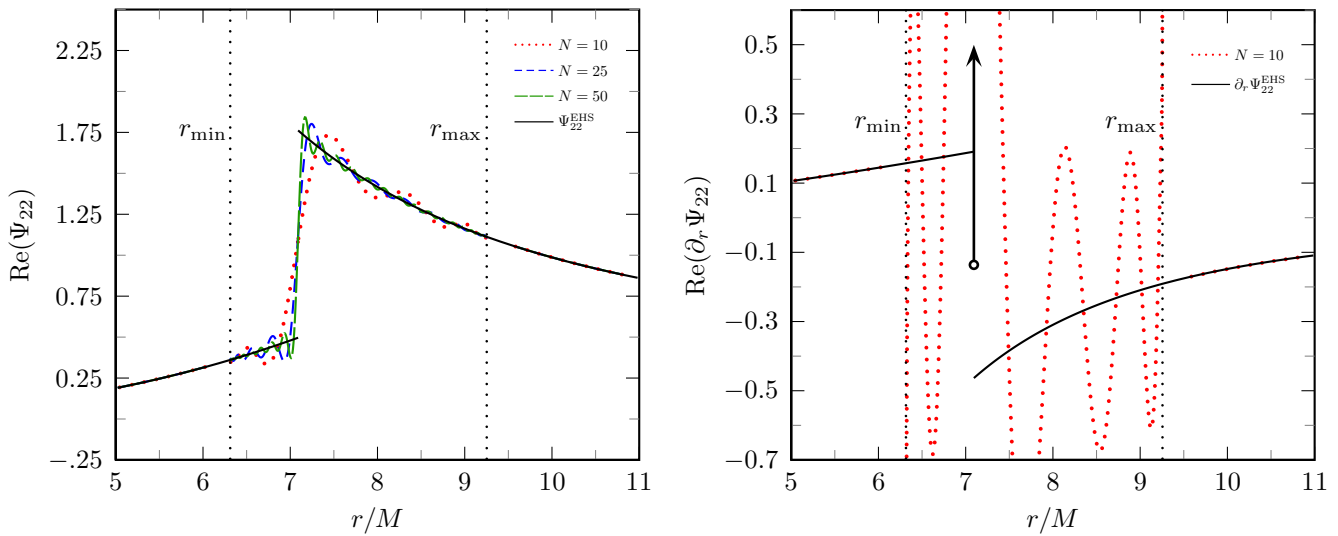}
		\caption
		{
			\label{fig:jumpStd}
			The standard FD approach to 
			reconstructing the TD master function and its $r$ derivative.
			The left panel shows
			$\Psi^{\rm std}_{22}$ and the right shows
			$\pa_r \Psi^{\rm std}_{22}$ at $t = 51.78 M$ for
			a particle orbiting with $p = 7.50478$ and 
			$e = 0.188917$.  This figure is 
			analogous to FIG.~1 of BOS \cite{BOS}.  
			Partial sums are computed with Eq.~(\ref{eq:PsiStd})
			and shown for different $N$.
			For 
			contrast we plot the converged solution from the new 
			method with a solid curve (see FIG. \ref{fig:jumpEHS}).
			The arrow in the right panel gives a notional 
			representation of the delta function singularity 
			present in $\pa_{r} \Psi_{22}$; the amplitude of 
			this singular term is related to the jump in 
			$\Psi_{22}$ seen in the left panel.
		}
	}
	\end{center}
}
\end{figure}

\begin{figure}[h!]
{
	\begin{center}
	{
		\includegraphics{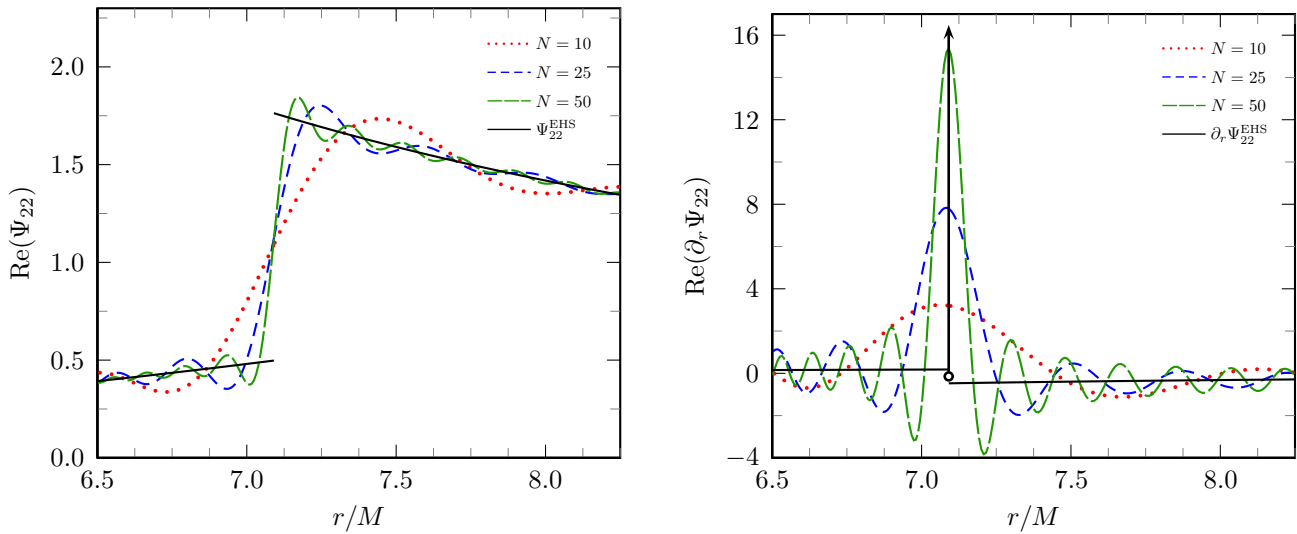}
		\caption
		{
			\label{fig:jumpStd2}
			An alternate view of the behavior presented 
			in FIG.~\ref{fig:jumpStd}.
			A change in the scale in the left panel emphasizes 
			the Gibbs overshoots in $\Psi_{22}$.  On the right,
			a zoom-out of the vertical scale more clearly indicates
			the attempt of the Fourier synthesis to capture the
			delta function at $r_{p}(t)$.
		}
	}
	\end{center}
}
\end{figure}

Looking ahead somewhat, we use our numerical code to obtain a particular 
spherical harmonic amplitude, $\Psi_{22}(t,r)$ ($\ell=2$, $m=2$), and its 
radial derivative, $\pa_r \Psi_{22}(t,r)$.  We can also use the code to 
assemble the standard partial Fourier sums (see FIGs.~\ref{fig:jumpStd} and
\ref{fig:jumpStd2}).  
\emph{We find that the Gibbs problem with the standard approach is 
significantly worse in the gravitational case (in Regge-Wheeler gauge) than 
it is for the scalar field.}  In the present case the field 
itself has a discontinuity and the radial derivative is both discontinuous 
as $r\rightarrow r_p(t)$ and also has a delta function singularity at 
$r_p(t)$.  The left panels of 
FIGs.~\ref{fig:jumpStd} and
\ref{fig:jumpStd2} are familiar; the partial sums have 
difficulty representing the jump discontinuity and overshoot the exact 
solution (solid curve).  In the right panels, the singularity at $r_p(t)$ 
wreaks havoc on the ability of the Fourier synthesis to represent the 
exact solution.

On a bright note, outside the range of the source, the standard solution 
converges exponentially fast.  Nevertheless, in the source region 
between $r_{\rm min}$ and $r_{\rm max}$ the convergence will be algebraic 
in general and disastrous at the location of the particle.  A discontinuous 
(or worse, singular) function cannot be accurately represented by a sum of 
smooth functions.

We now generalize the EHS method to the gravitational case.  We start by 
recognizing that 
$R^{\pm}_{\ell mn}$ from Eq.~(\ref{eq:freqHomogSol}) are valid solutions to
the homogeneous version of Eq.~(\ref{eq:FDwaveEq}) throughout 
the entire domain outside the black hole,
\be
\label{eq:FD_EHS}
R^\pm_{\ell mn} (r) 
= C^{\pm}_{\ell mn} \hat R_{\ell mn}^\pm (r), \q \q r > 2M.
\ee
Next, we use these to define the \emph{time-domain extended homogeneous 
solutions},
\be
\label{eq:TD_EHS}
\Psi^\pm_{\ell m} (t,r) 
\equiv \sum_n R^\pm_{\ell mn} (r) \, e^{-i \o_{mn} t}, \q \q r > 2M,
\ee
which result from inserting $R^\pm_{\ell mn}$ into 
Eq.~(\ref{eq:FDPsi}).
The central claim is then that for any $t$ and $r$ the actual solution to 
the inhomogeneous wave equation (\ref{eq:TDwaveEq}) is given by 
\be
\label{eq:weaksolution}
\Psi_{\ell m}(t,r) = \Psi^{\rm EHS}_{\ell m}(t,r) \equiv
\Psi_{\ell m}^+ (t,r) \, \th \left[ r - r_p(t) \right] +
\Psi_{\ell m}^- (t,r) \, \th \left[ r_p(t) - r \right].
\ee
The argument made by BOS can be extended to the gravitational case and goes 
as follows:
\begin{itemize}
\item We denote the desired true solution of the inhomogeneous wave equation 
as $\Psi_{\ell m}$.  Outside the domain of the source 
($r < r_{\rm min}, r_{\rm max} < r$) 
$\Psi_{\ell m} = \Psi^{\rm std}_{\ell m} = \Psi^{\rm EHS}_{\ell m}$ because 
there  $R_{\ell mn} = R^\pm_{\ell mn}$.
\item It is assumed that $\Psi_{\ell m} (t,r)$ is analytic in the entirety 
of the two regions $2M < r < r_p(t)$ and $r_p (t) < r$ (excluding only a 
neighborhood of $r_p(t)$).
\item Because the homogeneous solutions $\Psi^{\pm}_{\ell m}$ are expected 
to be analytic everywhere, $\Psi^{\rm EHS}_{\ell m} (t,r)$ will be analytic 
in the two regions discussed above (excluding only a neighborhood of 
$r_p(t)$).  (See the extended discussion BOS have about this.)
\item Because $\Psi_{\ell m}$ and $\Psi^{\rm EHS}_{\ell m}$ are identical 
outside the region of libration, and they are both analytic everywhere up 
to the location of the source, they must be equal over that entire domain.
\end{itemize}

\begin{figure}[h!]
{
	\begin{center}
   {
		\includegraphics{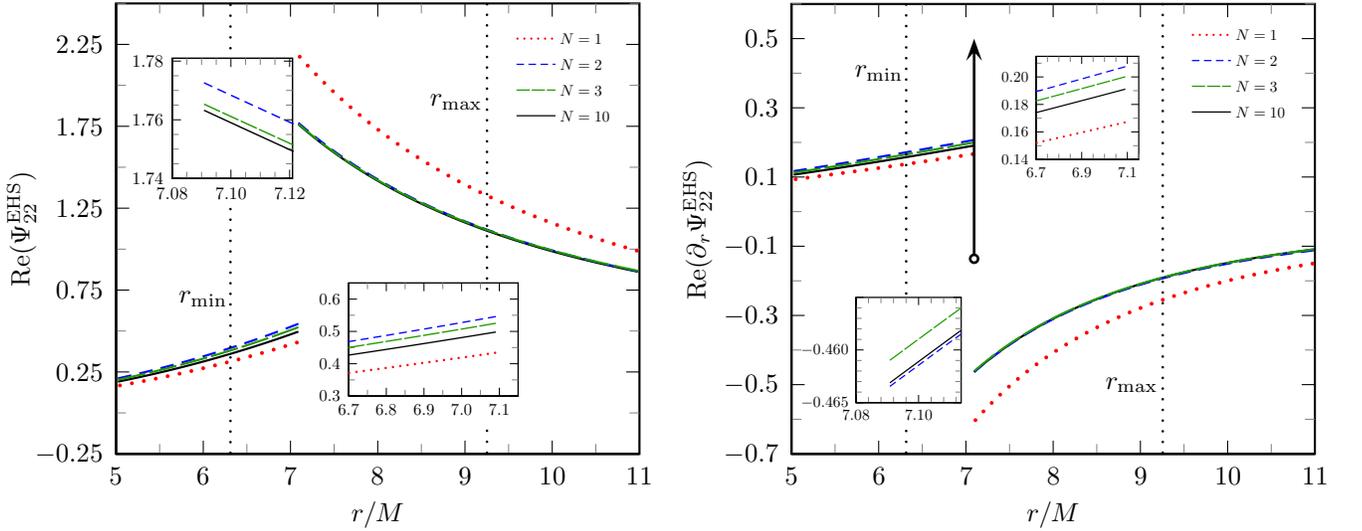}
		\caption
		{
			\label{fig:jumpEHS}
			The EHS approach 
			to reconstructing the TD master function 
			and its radial derivative.
			As in FIG.~\ref{fig:jumpStd}, we give
			$\Psi^{\rm EHS}_{22}$ and 
			$\pa_{r} \Psi^{\rm EHS}_{22}$ at $t = 51.78 M$ for a
			particle orbiting with $p = 7.50478$ and 
			$e = 0.188917$.  Partial sums of 
			$\Psi^{\rm EHS}_{22}$ are computed
			from Eq.~(\ref{eq:TD_EHS}), with a range of $-N \le n \le N$.  
			The full $\Psi^{\rm EHS}_{22}$ 
			and its $r$ derivative result from $N=10$, which
			gives agreement in the jumps 
			in $\Psi^{\rm EHS}_{22}$ and
			$\pa_{r} \Psi^{\rm EHS}_{22}$
			to a relative error of $10^{-10}.$  
			On the right, the presence of
			a delta function singularity is notionally depicted
			with an arrow.  
			The time dependent amplitude of this singularity is
			separately computable from the jump in $\Psi_{22}$.
		}	
	}
	\end{center}
}
\end{figure}

Here we provide an additional justification for the assumed form of the 
solution given in Eq.~(\ref{eq:weaksolution}).  The source term of the 
wave equation is a distribution, or generalized function 
\cite{Lighthill_1958}.  Accordingly, any solution of Eq.~(\ref{eq:TDwaveEq}) 
will be a weak solution--a generalized function itself--with loss of 
(classic) differentiability at the singular point $r_p(t)$.  To determine 
the suitability of Eq.~(\ref{eq:weaksolution}) as a solution of 
Eq.~(\ref{eq:TDwaveEq}), we generalize the concept of differentiation to 
encompass distributions.  Thus, for example, $d\th(z)/dz = \d (z)$.  We can 
then take Eq.~(\ref{eq:weaksolution}) as an ansatz, substitute in 
Eq.~(\ref{eq:TDwaveEq}), and determine what conditions are required that it be
a (weak) solution.  For clarity, in the rest of this section we suppress 
the $\ell$ and $m$ indices.

Rather than use the RWZ equation as it stands, we introduce a coordinate 
transformation to fix the position of the singularity.  Defining
$z \equiv r - r_p(t), \bar t \equiv t$, the derivatives transform as
$\pa_{r_*} = f(r) \pa_z$ and  $\pa_t = \pa_{\bar t} - \dot r_p \pa_z$,
and the wave equation (\ref{eq:TDwaveEq}) becomes
\be
L(\Psi) = - \pa_{\bar t}^2 \Psi  + \l f^2 - {\dot r_p}^2  \r 
\pa_z^2 \Psi  + 2 \dot r_p \pa_{\bar t} \pa_z \Psi  
+ \Big( \ddot r_p  + \l f \pa_z  f \r \Big) \pa_z \Psi
- V   \Psi
= \tilde G \, \d (z)
+ \tilde F \, \d' (z).
\label{eq:comovingWave}
\ee
Now we assume that $\Psi$ has the form given in Eq.~(\ref{eq:weaksolution}) 
and substitute it into Eq.~(\ref{eq:comovingWave}).  The functions $\Psi^+$ 
and $\Psi^-$ are 
differentiable and satisfy the homogeneous equation, $L(\Psi^{\pm})= 0$.  
A term of the form $L(\Psi^{+})\, \th(z) + L(\Psi^{-})\, \th(-z)$ appears 
in (\ref{eq:comovingWave}) and drops out.  Other singular terms remain, 
created by derivatives of the Heaviside function, and we are left with
\be
\label{eq:InitialSingular}
\l f^2 - {\dot r_p}^2  \r 
\Big( \llbracket \pa_r \Psi \rrbracket_{p}  \, \d (z) 
+ \llbracket \Psi \rrbracket_{p}  \, \d' (z) \Big)
+ 2 \dot r_p \pa_{\bar t} 
\Big( \llbracket  \Psi \rrbracket_{p}  \, \d (z) \Big) 
+ \Big( \ddot r_p  + \l f \pa_z  f \r \Big)  
\llbracket \Psi \rrbracket_{p}  \, \d (z)
= \tilde G \, \d ( z )
+ \tilde F \, \d' ( z ) .
\ee
where
\be
\label{eq:PsiJumpDef}
\llbracket \Psi \rrbracket_{p} (t)
\equiv \Psi^+ 
\left( t, r_p(t) \right) - \Psi^- \left( t, r_p(t) \right), 
\q \q
\llbracket \pa_r \Psi\rrbracket_{p} (t)
\equiv \pa_r \Psi^+
 \left( t, r_p(t) \right) 
 - \pa_r \Psi^- \left( t, r_p(t) \right)
\ee
are the jumps in $\Psi$ and $\pa_r \Psi$ at $z = 0$.
Na\"ively, we might expect that we can simply equate the coefficients 
of $\d$ on the two sides of Eq.~\ref{eq:InitialSingular}, while doing 
the same with the $\d'$ coefficients.  However, the $\d'$ term on the left 
hand side must first be fully evaluated (as a function of time) at the 
location of the particle.  To do this, we use the identities in 
Eqs.~(\ref{eq:deltaEvalR}) and (\ref{eq:deltaPrimeEvalR}), which leaves
\begin{multline}
\l f_{p}^2 - {\dot r_p}^2  \r 
\llbracket  \pa_r \Psi \rrbracket_{p}  \, \d (z) 
+ \l f_{p}^2 - {\dot r_p}^2  \r 
\llbracket \Psi \rrbracket_{p} \, \d'(z) 
- 2 \l f_{p} \pa_z f_{p} \r \llbracket \Psi \rrbracket_{p} \, \d (z) 
+ 2 \dot r_p \pa_{\bar t} 
\Big( \llbracket \Psi \rrbracket_{p}  \Big)  \,  \d (z) \\
+ \Big( \ddot r_p  + \l f_{p} \pa_z  f_{p} \r \Big) 
\llbracket \Psi \rrbracket_{p}  \, \d (z) 
= \tilde G \, \d ( z )  
+ \tilde F \, \d' ( z ) ,
\end{multline}
where $f_p \equiv f(r_p(t))$.  Note that there is no comparable expansion 
on the right side from the $\tilde F \, \d' (z)$ term because $\tilde F$ is 
already fully evaluated at $r=r_{p}(t)$, by design.  From here, we read off 
the jumps in $\Psi$ and its $r$ derivative at $r_p(t)$ from the coefficients 
of $\d'$ and $\d$, respectively.  Returning to Schwarzschild coordinates 
and using Eqs.~(\ref{eq:removeRDot}) to remove $\ddot r_p$ and 
$\dot r_p^2$ terms, we find
\begin{align}
\label{eq:psiJump}
\llbracket \Psi \rrbracket_{p} (t)
&= \frac{{\cal{E}}^2}{f_{p}^2 U_{p}^2} \tilde F(t), 
& \llbracket \pa_r \Psi \rrbracket_{p} (t)
&= \frac{{\cal{E}}^{2}}{ f_{p}^2 U_{p}^2}
\left[ \tilde G(t)
+ \frac{1}{U_{p}^{2} r_{p}^{2}}
\l 3 M -  \frac{{\cal{L}}^2}{r_{p}} + \frac{5 M{\cal{L}}^2}{r_{p}^2} \r 
\tilde F (t)  
- 2 \dot r_p \frac{d}{dt} 
\Big( \llbracket \Psi \rrbracket_{p}  \Big) \right].
\end{align}
From the standpoint of the original coordinates, the partial time derivative 
$\pa_{\bar t}$ becomes the convective, or total, time derivative along the
particle worldline.  

These jump conditions amount to internal boundary conditions that are 
necessary conditions on a solution to the inhomogeneous wave equation in 
the TD.  They were discussed by Sopuerta and Laguna \cite{SL_2008} and 
also later, with corrections, by Field et al.~\cite{FHL_2009}.  In our
FD-based calculations, they provide a powerful check on our transformation
of the solutions back to the TD.  Given the indirect way in which the 
Fourier transform of the source $S_{\ell m}$ determines the Fourier 
coefficients of the extended homogeneous solutions, considerable credence 
is lent to the method in seeing the partial sums of $\Psi^{\rm EHS}_{\ell m}$ 
converge toward satisfying these jump conditions.  Secondarily, the jump 
conditions provide useful stopping criteria in the numerical method (see 
Sec.~\ref{codeValidation}).

While not a focus of this paper, we consider briefly TD simulations.  There, 
to find a unique solution the internal boundary conditions must be augmented 
with initial data on a Cauchy surface and, potentially, outer boundary 
conditions.  Care must be exercised to switch on the source smoothly in 
the (near) future of the initial value surface \cite{FHL_2009} (also Lau, 
private communication).  Additionally, imposed initial data will not 
typically match long term periodic behavior induced by the source, and 
transients will sweep through the system for several dynamical times.  In 
contrast, in the FD approach, the proper outgoing and downgoing behavior 
at the outer boundaries is built in from the outset and only the steady state, 
periodic behavior is obtained.

\subsection{Computing normalization coefficients 
in the gravitational case}
\label{CoeffSec}

Finally, we provide some details on how the singular source is integrated 
to provide the matching normalization coefficients $C^+_{\ell mn}$ and 
$C^-_{\ell mn}$ that are used in Eq.~(\ref{eq:FD_EHS}).  BOS detail the 
calculation of normalization coefficients for the scalar monopole in 
their App.~C.  The gravitational case follows the same general idea, but 
involves some technical differences and challenges.  We start by combining 
Eqs.~(\ref{eq:CNorm}) and (\ref{eq:cPM}), giving
\be
C_{\ell mn}^{\pm} 
= \frac{1}{W_{\ell mn}} \int_{r_{\rm min}}^{r_{\rm max}} dr
 \ \frac{\hat R^{\mp}_{\ell mn} (r) Z_{\ell mn} (r)}{ f(r)}.
\ee
The FD source term $Z_{\ell mn} (r)$ comes from 
plugging Eq.~(\ref{eq:gravitySource}) into 
Eq.~(\ref{eq:FDSource}), yielding
\be
\label{eq:FDSource2}
Z_{\ell mn} (r) = \frac{1}{T_r} \int_0^{T_r} dt 
\ \Big( \tilde G_{\ell m} (t) \, \d [r - r_p(t)] + 
\tilde F_{\ell m} (t) \, \d' [r - r_p(t)] \Big) 
 e^{i \o_{mn} t} .
\ee
The equivalent integral BOS present for the scalar monopole is 
their Eq.~(C2), which they evaluate immediately by changing the 
integration variable from $t$ to $r_p$.  Here, with a derivative-of-the-delta 
function present (in RWZ gauge), the immediate evaluation of this 
integral produces terms that are singular at the turning points 
($\dot r_{p} = 0$).  These terms are no problem analytically, but they 
are troublesome when performing the final numerical integration.  
We therefore find it is advantageous to delay this integration.  
Plugging our expression for $Z_{\ell mn}$ in above, we have
\be
\label{eq:Cpm2}
C_{\ell mn}^\pm  = \frac{1}{W_{\ell mn} T_r} 
\int_{r_{\rm min}}^{r_{\rm max}} dr \ \frac{\hat R^\mp_{\ell mn} (r)}{f(r)}
\int_0^{T_r} dt \   
\Big( \tilde G_{\ell m} (t) \,  \d [r - r_p(t)] 	
+ \tilde F_{\ell m} (t) \, \d' [r - r_p(t)] \Big)
 e^{i \o_{mn} t}  .
\ee
In order to avoid the singularity at the turning points, we switch 
the order of integration.  The integration of the delta function itself is 
then straightforward.  The derivative of $\d$ term requires an integration 
by parts.  Because of the compact support of the source term, we can extend 
the range of integration and no surface terms appear.  We are left with
\be
C_{\ell mn}^\pm  
=   \frac{1}{W_{\ell mn} T_r} \int_0^{T_r}
\Bigg[ 
 \frac{1}{f_{p}} \hat R^\mp_{\ell mn} (r_{p})
 \tilde G_{\ell m} (t)
+ \l \frac{2M}{r_{p}^2 f_{p}^{2}} \hat R^\mp_{\ell mn} (r_{p})
 - \frac{1}{f_{p}} 
 \frac{d \hat R^\mp_{\ell mn} (r_{p})}{dr} \r \tilde F_{\ell m} (t)
 \Bigg]  e^{i \o_{mn} t}  \, dt,
\label{eq:EHSC}
\ee
where we use a $p$ subscript to indicate evaluation of a quantity at 
$r=r_{p}(t)$.  Our final integral is analogous to Eq.~(C7) in BOS.  

Here is a summary of key details of the application of the method in the 
gravitational case:
\begin{itemize}
\item The EHS method, applied to the 
gravitational case, gives exponentially converging solutions to 
Eq.~(\ref{eq:TDwaveEq}) everywhere, including the location of the particle.  
(See FIG.~\ref{fig:converge}.)
\item Working in Regge-Wheeler gauge, 
the gravitational TD source term contains a delta 
function and a derivative-of-the-delta function, which cause 
$\Psi_{\ell m}$ to exhibit a jump and $\pa_r \Psi_{\ell m}$ to exhibit
both a jump and a delta function singularity at 
the particle's location. 
(See FIG.~\ref{fig:jumpEHS}.)  In the 
scalar case, the field is piecewise continuous at the particle, 
with a jump in the $r$ derivative. (See FIG.~3 in BOS.)   
\item Eq.~(\ref{eq:EHSC}) is valid for all radiative multipoles 
($\ell \ge 2$).  The $\ell = 0, 1$, modes must be handled separately.
\item Martel's \cite{Martel_2004}
$G_{\ell m} (t,r)$ and $F_{\ell m} (t,r)$ 
from Eq.~(\ref{eq:SMartelForm}) are not in fully evaluated form.  
As discussed in App.~\ref{generalized}, for a given multipole, 
unique functions of time
$\tilde F_{\ell m}(t) \equiv F_{\ell m} \left( t, r_p(t) \right)$ and 
$\tilde G_{\ell m}(t) 
\equiv G_{\ell m} \left( t, r_p(t) \right) - \pa_r F_{\ell m} 
\left( t, r_p(t) \right)$ emerge after fully applying the delta function 
constraint.  We use the tilde to distinguish fully evaluated coefficients.
\item 
In practice, we take advantage of the
fact that some of the functions in the integrand of Eq.~(\ref{eq:EHSC})
are even over the period of radial libration, while others are odd.
Then, rather than integrating over $t$ from $0 \to T_r$, we can limit the
range of integration to $0 \to T_r / 2$.  Further, we change variables to 
$\chi$, as shown in Sec.~\ref{orbits} and integrate from $0 \to \pi$.
\item For $\Psi_{\ell m}^{\rm even}$ we use the Zerilli-Moncrief 
master function, and for   $\Psi_{\ell m}^{\rm odd}$ we use the 
Cunningham-Price-Moncrief master function.  This formulation works for 
any master function that obeys a Regge-Wheeler-like equation and has a 
source term that can be written in the form of 
Eq.~(\ref{eq:gravitySource}).
\end{itemize}

\section{Numerical method and results from mode integrations}
\label{modeIntegration}

\subsection{Algorithmic roadmap}
\label{method}

Here, we explain the specific steps  our code 
takes to solve the inhomogeneous 
wave equation (\ref{eq:TDwaveEq}).  
There are several stages to the process, and 
at each step we compute at least one more order of magnitude accuracy 
than is needed at the subsequent step.
The code 
is written in C, and we use the Numerical Recipes adaptive step size 
fourth order Runge-Kutta integrator \cite{NRC_1993}.  

\begin{enumerate}
\item Specify an orbit through a choice of the semi-latus rectum $p$ 
and eccentricity $e$.
\item Numerically integrate Eqs.~(\ref{eq:O_r}) and (\ref{eq:O_phi}) 
to get the fundamental frequencies of the system, $\O_r$ and 
$\O_\varphi$, and hence $\o_{mn} = m \O_{\varphi} + n \O_r $.  
\item
\label{chooseLM}
 Choose a specific $\ell$ and $m$.  If $\ell + m$ is even (odd), 
use even (odd) parity potential and source terms.   Choose starting $n$. 
(See Sec.~\ref{codeValidation}.)
\item 
\label{SourceFreeSol}
Solve the homogeneous version of Eq.~(\ref{eq:FDwaveEq}) to
get unit normalized radial mode functions, $\hat R_{\ell m n}^{\pm}$, in the 
source-free region:
\begin{itemize}
\item Use the asymptotic expansion (see App.~\ref{asympExp}) to 
set an ``up'' plane  wave boundary condition at $r_* \to + \infty$, 
as in Eq.~(\ref{eq:up}).  Numerically integrate up to 
the region of the source at $r_*^{\rm max}$ to get 
$\hat R^+_{\ell mn}$.  (We let $r_*^{\rm min / max}$ be
the $r_*$ value corresponding to $r_{\rm min / max}$.)
\item Use a convergent Taylor expansion to set an ``in'' 
plane wave boundary condition (Eq.~(\ref{eq:in})) at modestly 
negative $r_{*}$.
Numerically integrate up to the region of the source 
at $r_*^{\rm min}$ to get $\hat R^-_{\ell mn}$.
\end{itemize}
\item Solve the homogeneous version of Eq.~(\ref{eq:FDwaveEq}) 
to continue the unit normalized radial mode functions, 
$\hat R_{\ell m n}^{\pm}$, into the source region, while also 
computing the normalization coefficients $C_{\ell m n}^{\pm}$:
\begin{itemize}
\item Simultaneously integrate Eqs.~(\ref{eq:FDwaveEq}) and (\ref{eq:EHSC})
from $\chi = 0 \to \pi$ (equivalently $t = 0 \to T_{r}/2$ and 
$r = r_{\rm min} \to r_{\rm max}$).  This gives $\hat R_{\ell m n}^{-}$ in the
region of the source and $C_{\ell m n}^{+}$.
\item Simultaneously integrate Eqs.~(\ref{eq:FDwaveEq}) and (\ref{eq:EHSC})
from $\chi = -\pi \to 0$ (equivalently $t = -T_{r}/2 \to 0$ and 
$r = r_{\rm max} \to r_{\rm min}$).  This gives $\hat R_{\ell m n}^{-}$ in the
region of the source and $C_{\ell m n}^{-}$.
\end{itemize}
As discussed in Sec.~\ref{CoeffSec}, 
the integrand in Eq.~(\ref{eq:EHSC}) contains parts which are even and parts 
which are odd over the radial period. By keeping the correct terms, 
we can get away with efficiently integrating over only half the period.
\item Use the coefficients to normalize the homogeneous solutions 
outside \emph{and inside} the range of the source, as in Eq.~(\ref{eq:FD_EHS}).
\item Assess whether there is convergence of the partial sum over $n$. 
(Again, see Sec.~\ref{codeValidation}.)
\begin{itemize}
\item If yes, we are finished with this $\ell, m$ mode.  
\item If no, return to Step \ref{SourceFreeSol} with the next $n$.
\end{itemize}
\end{enumerate}

\subsection{Energy and angular momentum fluxes at $r_* = \pm\infty$}

To evaluate the energy and angular momentum fluxes at $r_{*} = \pm \infty$ 
we use the Isaacson stress-energy tensor.  The energy and angular 
momentum fluxes, for each $\ell, m$ mode, can be written as \cite{Thorne_1980}
\be
\label{eq:EAndLDot}
\dot E^\pm_{\ell m} = \frac{1}{64 \pi} 
\frac{(\ell+2)!}{(\ell-2)!}
\left| \dot \Psi^{\pm}_{\ell m} (t, r) \right|^2, 
\q \q
\dot L^\pm_{\ell m} = 
\frac{ i m }{64 \pi} \frac{(\ell+2)!}{(\ell-2)!} 
 \dot \Psi^{\pm}_{\ell m} (t, r)  \Psi^{\pm \, *}_{\ell m} (t, r).
\ee
Here, an asterisk signifies complex conjugation. 
(We use $\Psi_{\ell m}^{\rm even}$ when $\ell + m$ is even and 
$\Psi_{\ell m}^{\rm odd}$ when $\ell + m$ is odd.  In general there would be
contributions from both $\Psi_{\ell m}^{\rm even}$ 
and $\Psi_{\ell m}^{\rm odd}$ for 
each mode, but our choice of $\th_{p} = \pi/2$ leads to one of these 
functions vanishing for each $\ell$ and $m$ combination.)
In terms of FD amplitudes the expressions become 
\begin{align}
\begin{split}
\dot E^\pm_{\ell m} &= \frac{1}{64 \pi} 
\frac{(\ell+2)!}{(\ell-2)!}
 \sum_{n, n'} \o_{mn} \o_{mn'} R_{\ell mn}^{\pm}  
R^{\pm \, *}_{\ell mn'}  
 e^{-i \l \o_{mn} - \o_{mn'} \r t}, \\
 \dot L^\pm_{\ell m} &= 
\frac{ m }{64 \pi} \frac{(\ell+2)!}{(\ell-2)!} 
 \sum_{n, n'} \o_{mn} R_{\ell mn}^{\pm}  
R^{\pm \, *}_{\ell mn'}  
 e^{-i \l \o_{mn} - \o_{mn'} \r t} .
\end{split}
\end{align}
As is well known, the fluxes must be suitably averaged over time or space 
to obtain meaningful, invariant results.  We average these quantities 
in time over one radial oscillation, which yields 
\be
\left\langle \dot E^\pm_{\ell m} \right\rangle 
= \frac{1}{64 \pi} 
\frac{(\ell+2)!}{(\ell-2)!}
\sum_{n} \o_{mn}^{2} \left| C^{\pm}_{\ell mn} \hat R^\pm_{\ell mn}\right|^2,
\q \q
\left\langle \dot L^\pm_{\ell m} \right\rangle = 
\frac{m}{64 \pi} \frac{(\ell+2)!}{(\ell-2)!} 
\sum_{n} \o_{mn} \left| C^{\pm}_{\ell mn} \hat R^\pm_{\ell mn} \right|^2 .
\ee
Here, we have also introduced 
$R^{\pm}_{\ell mn} = C^{\pm}_{\ell mn} \hat R^{\pm}_{\ell mn}$.  As discussed 
in App.~\ref{asympExp}, we can write the radial function as 
$\hat R^\pm_{\ell mn} (r) = J^\pm_{\ell mn} (r) e^{\pm i \o_{mn} r_*}$, where 
$J^\pm_{\ell mn} (r) \to 1$ as $r_* \to \pm \infty$.  Therefore, 
if we set $J_{\ell mn}^\pm = 1$, we can 
evaluate the fluxes at $r_* = \pm \infty$, leaving
\begin{align}
\label{eq:E&LDotInf}
\left\langle \dot E^{\pm \infty}_{\ell m} \right\rangle 
= \frac{1}{64 \pi} 
\frac{(\ell+2)!}{(\ell-2)!}
\sum_{n} \o_{mn}^{2} \left| C^{\pm}_{\ell mn} \right|^2, 
\q \q
\left\langle \dot L^{\pm \infty}_{\ell m} \right\rangle = 
\frac{m}{64 \pi} \frac{(\ell+2)!}{(\ell-2)!} 
\sum_{n} \o_{mn} \left| C^{\pm}_{\ell mn} \right|^2 .
\end{align}

\subsection{Code validation}
\label{codeValidation}

To compute the total energy and angular momentum fluxes, we must sum 
Eqs.~(\ref{eq:E&LDotInf}) over $\ell$ and $m$.
The resulting expressions are formally over the ranges 
$ 2 \le \ell \le \infty$, $-\ell \le m \le \ell$, 
$ -\infty \le n \le \infty$.  When computing $\dot E$ and $\dot L$ 
numerically, we put limits on each of these sums.  
To begin with, the low $\ell$ modes matter more than the high ones.
But, the more eccentric an orbit, the 
more $\ell$'s must be computed to achieve 
the same precision in our final values.
For the orbits we considered in Table 
\ref{totalFluxTable}, in order to achieve a relative 
precision of $10^{-12}$ in our final flux values, 
the highest $\ell$ necessary was $\ell = 29$.  (See Sec.~\ref{results}.)

Because of the 
symmetry of the spherical harmonics, the fluxes
from any given $-m$ mode are equal to those from the corresponding 
$+m$ mode.  Therefore, we fold the 
negative $m$ modes over onto the positive ones, and simply multiply
each positive $m$ mode by two.  
Additionally, 
as $\ell$ gets larger, it is no longer necessary to compute all
$m$ values.  As can be seen in Table \ref{highEFluxTable}, 
for a given $\ell$, the largest 
$\dot E_{\ell m}^{\infty / H}$ 
and $\dot L_{\ell m}^{\infty / H}$ contributions come from
the $m = \ell$ mode.  We start at $m = \ell$ and decrement $m$ until
the fluxes are no longer significant.  For low $\ell$ values we still
wind up computing all $0 \le m \le \ell$, 
but as $\ell$ increases, we need progressively fewer $m$ modes.  

Determining the necessary $n$'s is a bit more involved. 
For a given $\ell$ and $m$, there is a range, $n_{\rm min}$ to 
$n_{\rm max}$, over which we sum in order to achieve our desired precision.
Looking at Table \ref{highEFluxTable}, it is evident that when $m = 0$, 
the range of $n$ is essentially centered on 0.  For these modes,
we start with $n = 0$, and compute fluxes for all positive modes.
When we have seen no change
to any of the flux values 
(at a pre-specified level of precision) for several consecutive modes,
we stop and repeat the process for the negative $n$'s.
As $m$ increases, this range of $n$'s shifts
more and more into the positive.
For any $\ell$, the $m = \ell$ mode has far more positive $n$ modes
than negative.  Eventually, $\ell$ becomes so large that $n_{\rm min} > 0$
for the $m = \ell$ mode.  For modes where we suspect that $n_{\rm min} > 0$, 
we find it advantageous to start with a rough sweep of a large range of 
possible $n$ values.  We calculate $\dot E_{\ell m n}^{\infty}$  (the energy
flux at $r = +\infty$ from one $n$ mode) to low precision for a small
number of $n$, spaced out over this range.  The $n$ for which 
we find the largest $\dot E_{\ell m n}$ will be near the center of 
the $n_{\rm min}$ to $n_{\rm max}$ range.
We then perform our high precision mode 
integrations for all significant $n$ values above and below this $n$.

If we are interested in a local calculation (as one would perform for a 
SF evaluation), we have a different method for determining which $n$'s 
are significant.   We still use the energy fluxes to find the approximate 
center of the significant $n$ range, but for the ``breaking condition'' we 
compute $n$'s until the jumps in $\Psi_{\ell m}$ and 
$\pa_{r} \Psi_{\ell m}$ converge properly, as follows:
\begin{itemize}
\item Use Eq.~(\ref{eq:TD_EHS}) to compute a partial mode sum approximation
of both 
$ \Psi_{\ell m}^\pm (t, r_p)$ and 
$\pa_r  \Psi_{\ell m}^\pm (t, r_p)$ for a large number of times 
$t_k$ throughout the orbit.  
\item Numerically evaluate the jumps in those partial sums
\begin{align}
\left\llbracket  \Psi^{N}_{\ell m} \right\rrbracket_{p}
\equiv  \Psi_{\ell m}^+ \l t, r_p \r
-  \Psi_{\ell m}^- \l t, r_p \r, 
\q \q
\left\llbracket \pa_r \Psi^{N}_{\ell m}\right\rrbracket_{p}  
\equiv \pa_r  \Psi_{\ell m}^+ \l t, r_p \r
- \pa_r  \Psi_{\ell m}^- \l t, r_p \r,
\end{align}
for those times $t_k$.  
\item Compute the analytical values of 
$ \left\llbracket \Psi^{A}_{\ell m} \right\rrbracket_{p} $ and 
$\left\llbracket \pa_r \Psi^{A}_{\ell m}\right\rrbracket_{p}$ 
derived in Sec.~\ref {jumpCond}  for those times $t_k$.
\item If $\left\llbracket  \Psi^{N}_{\ell m} \right\rrbracket_{p} = 
 \left\llbracket \Psi^{A}_{\ell m} \right\rrbracket_{p}$
and 
$\left\llbracket \pa_r \Psi^{N}_{\ell m}\right\rrbracket_{p}  =
\left\llbracket \pa_r \Psi^{A}_{\ell m}\right\rrbracket_{p}$
at all times $t_k$, 
to a chosen precision, we have computed enough $n$ modes.
\item Otherwise more $n$ modes are needed.  As in the flux computation case 
above, we perform the mode calculations for the $n$ values above our starting 
$n$, and once that partial sum has converged to our desired precision, we
solve for the $n$'s below our starting $n$ until the jump values agree.  
\end{itemize}

\subsection{Results}

One of our most important results is the exponential convergence of  
$\Psi^{\rm EHS}_{\ell m}$
and its $r$ derivative at the location of the particle.  FIG.~\ref{fig:jumpEHS}
shows a partial sum of these two quantities converging after only a few 
modes.  Compare this to FIGs.~\ref{fig:jumpStd} 
and \ref{fig:jumpStd2}, which shows the standard 
FD approach.  In particular, note in those figures 
the failure of the standard approach to 
accurately represent $\pa_{r} \Psi_{\ell m}$, even after a large number of
modes.  This function is particularly badly behaved in the standard approach
as smooth functions attempt to capture a delta function.

\begin{figure}[h!]
\centering
	\includegraphics{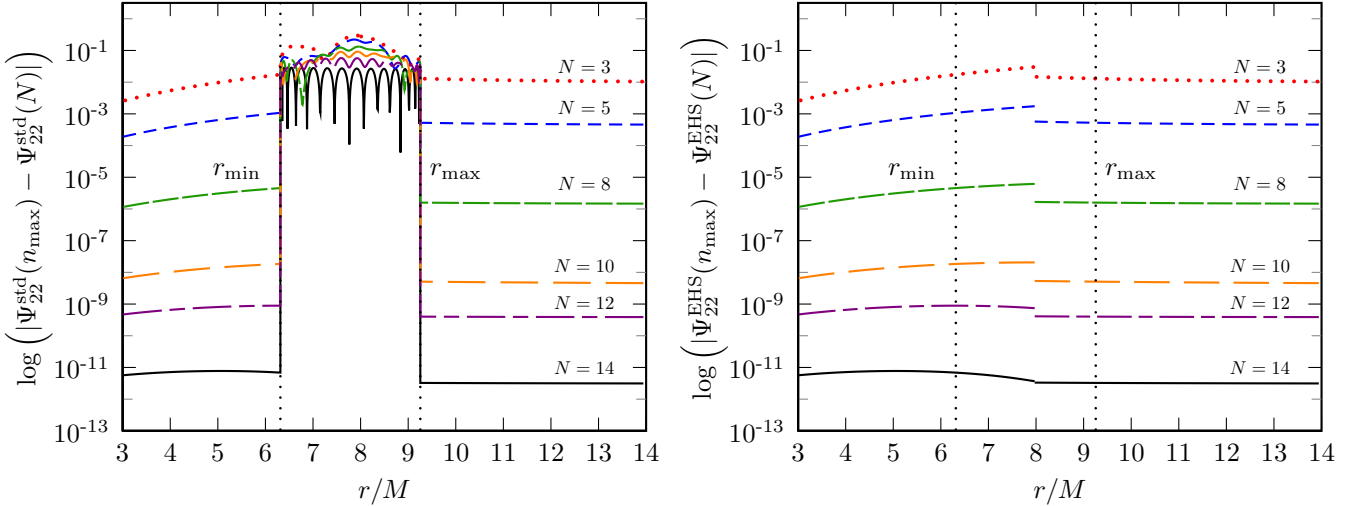}
	\caption
	{ 
		\label{fig:converge}
		A plot of the convergence of the master function 
		using the two methods.  For a particle orbiting
		with $p = 7.50478$ and $e = 0.188917$ at $t = 51.78 M$
		we compute the master function $\Psi_{22} (n_{\rm max})$  
		by summing over modes ranging from $-n_{\rm max} \le n \le n_{\rm max}$
		for $n_{\rm max} = 15$.
		We plot the log of the difference between $\Psi_{22} (n_{\rm max})$
		and the partial sum $\Psi_{22}(N)$, for different $N < n_{\rm max}$.
		For the standard approach (left), we see exponential convergence
		in the homogeneous region, but only algebraic convergence in the 
		region of the source.  The method of extended homogeneous 
		solutions (right) yields exponentially converging results at all 
		points outside \emph{and inside} the region of the source.
		The method of extended homogeneous solutions gives
		exponential convergence for $\pa_{r} \Psi^{\rm EHS}_{\ell m}$ as well.
	}
\end{figure}

Also of note is FIG.~\ref{fig:converge}, 
which shows that the convergence from the method of extended homogeneous 
solutions is indeed exponential, all the way up to the location of the 
particle.
Fast and accurate computation of $\Psi_{\ell m}$ and $\pa_{r} \Psi_{\ell m}$ 
at $r_p(t)$ will eventually be critical for reliable local SF calculations.

\label{results}

\begin{table*}[h!]
\begin{tabular}{ c || ll | ll | ll | ll}
\hline
\hline
& \multicolumn{2}{c|}{$\dot E^\infty_{\ell m} \ \ \l M^2 / \mu^2 \r$} 
& \multicolumn{2}{c|}{$\dot E^H_{\ell m} \ \ \l M^2 / \mu^2 \r$} 
& \multicolumn{2}{c|}{$\dot L^\infty_{\ell m}\ \ \l M / \mu^2 \r$} 
& \multicolumn{2}{c}{$\dot L^H_{\ell m}\ \ \l M / \mu^2 \r$}   \\
\hline
\hline
$p = 7.50478$, $e = 0.188917$  &&&&&&\\
This Paper, $\ell_{\rm max} = 23$ 
&  3.16899989185 & $\times 10^{-4}$ 
& 5.23247295625 & $\times 10^{-7}$  
& 5.96755215609 & $\times 10^{-3}$ 
& 8.71943028067 & $\times10^{-6} $ \\
{\rm Fujita et al.} 
& 3.16899989184 & $\times 10^{-4}$ 
&  \multicolumn{2}{c|}{N/A} 
& 5.96755215608 & $\times10^{-3}$ 
& \multicolumn{2}{c}{N/A} \\
\hline
$p = 8.75455$, $e = 0.764124$  &&&&&&\\
This Paper, $\ell_{\rm max} = 29$ 
&	2.12360313360 		& $\times 10^{-4}$ 
&	2.27177440621 		& $\times 10^{-6}$  
&	2.77735939025		& $\times 10^{-3}$ 
&	2.22781961809	 	& $\times 10^{-5}$ \\
{\rm Fujita et al.} 
& 2.12360313326 & $\times 10^{-4}$ 
&  \multicolumn{2}{c|}{N/A} 
& 2.77735938996 & $\times10^{-3}$ 
& \multicolumn{2}{c}{N/A} \\
\hline
\hline
\end{tabular}
\caption{
\label{totalFluxTable}
Total energy and angular momentum fluxes for eccentric 
orbits, compared with those from Fujita et al., published in \cite{BS_2010}.}
\end{table*}

In order to check our code's accuracy, we computed energy 
and angular momentum fluxes for
circular and eccentric orbits.  Our circular orbit fluxes agree, mode-by-mode, 
with published results (e.g. Cutler et al. \cite{CFPS_1993}) to high precision.  
For eccentric orbits, we are only aware that
total energy and angular momentum fluxes have been
published.  Our FD 
results agree with the fluxes at $r \to \infty$  of Fujita et al., 
published in \cite{BS_2010} to at least $10^{-9}$.  These are included in
Table \ref{totalFluxTable}.  Fujita et al. have also published horizon 
energy fluxes \cite{Fujita_2009}, 
which we agree with, to at least $10^{-9}$ for a range of 
eccentricities.  These are given in Table \ref{Fujita2}.

\begin{table*}
\begin{tabular}{ c || ll | ll | ll | ll}
\hline
\hline
& \multicolumn{2}{c|}{$\dot E^\infty_{\ell m} \ \ \l M^2 / \mu^2 \r$} 
& \multicolumn{2}{c|}{$\dot E^H_{\ell m} \ \ \l M^2 / \mu^2 \r$} 
& \multicolumn{2}{c|}{$\dot L^\infty_{\ell m}\ \ \l M / \mu^2 \r$} 
& \multicolumn{2}{c}{$\dot L^H_{\ell m}\ \ \l M / \mu^2 \r$}   \\
\hline
\hline
$p = 10$, $e = 0.1$  &&&&&&\\
This Paper
& 6.31752474718 & $\times 10^{-5}$ 
& 1.53365819446 & $\times 10^{-8}$  
& 1.95274165241 & $\times 10^{-3}$ 
& 4.48832141611 & $\times10^{-7} $ \\
{\rm Fujita et al.} 
& 6.31752474720 & $\times 10^{-5}$ 
& 1.53365819445 & $\times10^{-8}$ 
& \multicolumn{2}{c|}{N/A} 
& \multicolumn{2}{c}{N/A} \\
\hline
$p = 10$, $e = 0.5$  &&&&&&\\
This Paper
& 	9.27335011599 	& $\times 10^{-5}$ 
&	1.41298859263 	& $\times 10^{-7}$  
&	1.97465149446 	& $\times 10^{-3}$ 
&	2.15617302381 	& $\times10^{-6} $ \\
{\rm Fujita et al.} 
& 9.27335011503 & $\times 10^{-5}$ 
& 1.41298859260 & $\times10^{-7}$ 
& \multicolumn{2}{c|}{N/A} 
& \multicolumn{2}{c}{N/A} \\
\hline
$p = 10$, $e = 0.7$  &&&&&&\\
This Paper
& 9.46979134556 & $\times 10^{-5}$ 
& 3.55415030147  & $\times 10^{-7}$  
& 1.63064691133 & $\times 10^{-3}$ 
& 4.20771917244 & $\times10^{-6} $ \\
{\rm Fujita et al.} 
& 9.46979134409 & $\times 10^{-5}$ 
& 3.55415030114 & $\times10^{-7}$ 
& \multicolumn{2}{c|}{N/A} 
& \multicolumn{2}{c}{N/A} \\
\hline
$p = 10$, $e = 0.9$  &&&&&&\\
This Paper
& 4.194264692 & $\times 10^{-5}$ 
& 3.652142848  & $\times 10^{-7}$  
& 5.982866119 & $\times 10^{-3}$ 
& 3.518978461 & $\times10^{-6} $ \\
{\rm Fujita et al.} 
& 4.19426469206 & $\times 10^{-5}$ 
& 3.65214284306 & $\times10^{-7}$ 
& \multicolumn{2}{c|}{N/A} 
& \multicolumn{2}{c}{N/A} \\
\hline
\hline
\end{tabular}
\caption{
\label{Fujita2}
Energy and angular momentum fluxes for eccentric 
orbits, compared with those from Fujita et al.~\cite{Fujita_2009}.
Partial sums for all four orbits are truncated at $\ell_{\rm max} = 20$
for both papers.  Fujita et al.~obtained their numbers from integrating the 
Teukolsky equation. 
We include this table to show the agreement of our
horizon energy flux values. }
\end{table*}

For those wishing to reproduce our results, in Table 
\ref{highEFluxTable} we give mode-by-mode fluxes up to $\ell = 5$ at $r=\infty$ 
and down the black hole at $r = 2M$ for a particle in orbit with 
$p = 8.75455$ and $e = 0.764124$.  Included are the ranges of $n$ modes
summed over to achieve these results.

\begin{table*}[h!]
\begin{tabular}{cc || ll | ll | ll | ll | ll}
\hline
\hline
$\ell$ & $m$ 
& \multicolumn{2}{c|}{$\dot E^\infty_{\ell m} \ \ \l M^2 / \mu^2 \r$} 
& \multicolumn{2}{c|}{$\dot E^H_{\ell m} \ \ \l M^2 / \mu^2 \r$} 
& \multicolumn{2}{c|}{$\dot L^\infty_{\ell m}\ \ \l M / \mu^2 \r$} 
& \multicolumn{2}{c|}{$\dot L^H_{\ell m}\ \ \l M / \mu^2 \r$}  
& $n_{\rm min}$ & $n_{\rm max}$   \\
\hline
\hline
2 	 & 	 0 	 
& 	  1.27486196317 	 & 	 $\times 10^{-8}$ 	  
& 	  1.66171571270 	 & 	 $\times 10^{-8}$ 	  
&    \multicolumn{2}{c|}{0}
&    \multicolumn{2}{c|}{0}
& 	 -74 	 & 	 76 \\
 	 & 	 1 	 
& 	  1.15338054092 	 & 	 $\times 10^{-6}$ 	  
& 	  3.08063328605 	 & 	 $\times 10^{-7}$ 	  
& 	  1.44066000650 	 & 	 $\times 10^{-5}$ 	  
& 	  2.77518962557 	 & 	 $\times 10^{-6}$
& 	 -62 	 & 	 78 \\
 	 & 	 2 	 
& 	  1.55967717209 	 & 	 $\times 10^{-4}$ 	  
& 	  1.84497995136 	 & 	 $\times 10^{-6}$ 	  
& 	  2.07778922470 	 & 	 $\times 10^{-3}$ 	  
& 	  1.85014840343 	 & 	 $\times 10^{-5}$
& 	 -47 	 & 	 82 \\
\hline
3 	 & 	 0 	 
& 	  2.53527063853 	 & 	 $\times 10^{-11}$ 	  
& 	  1.23159713946 	 & 	 $\times 10^{-10}$ 	  
&    \multicolumn{2}{c|}{0}
&    \multicolumn{2}{c|}{0}
& 	 -84 	 & 	 85 \\
 	 & 	 1 	 
& 	  9.66848921204 	 & 	 $\times 10^{-10}$ 	  
& 	  2.47099909183 	 & 	 $\times 10^{-9}$ 	  
& 	  1.93528074730 	 & 	 $\times 10^{-8}$ 	  
& 	  2.10622579957 	 & 	 $\times 10^{-8}$
& 	 -66 	 & 	 87 \\
 	 & 	 2 	 
& 	  6.17859627641 	 & 	 $\times 10^{-7}$ 	  
& 	  1.29412677182 	 & 	 $\times 10^{-8}$ 	  
& 	  7.54192378736 	 & 	 $\times 10^{-6}$ 	  
& 	  1.23105502765 	 & 	 $\times 10^{-7}$
& 	 -48 	 & 	 93 \\
 	 & 	 3 	 
& 	  3.71507683858 	 & 	 $\times 10^{-5}$ 	  
& 	  8.07017762262 	 & 	 $\times 10^{-8}$ 	  
& 	  4.67102471030 	 & 	 $\times 10^{-4}$ 	  
& 	  7.99808068724 	 & 	 $\times 10^{-7}$
& 	 -34 	 & 	 99 \\
\hline
4 	 & 	 0 	 
& 	  1.14820411420 	 & 	 $\times 10^{-12}$ 	  
& 	  1.50591364139 	 & 	 $\times 10^{-12}$ 	  
&    \multicolumn{2}{c|}{0}
&    \multicolumn{2}{c|}{0}
& 	 -80 	 & 	 80 \\
 	 & 	 1 	 
& 	  4.58377338924 	 & 	 $\times 10^{-12}$ 	  
& 	  2.04365875527 	 & 	 $\times 10^{-11}$ 	  
& 	  4.50183584238 	 & 	 $\times 10^{-11}$ 	  
& 	  1.63314060565 	 & 	 $\times 10^{-10}$
& 	 -77 	 & 	 94 \\
 	 & 	 2 	 
& 	  1.59253324588 	 & 	 $\times 10^{-9}$ 	  
& 	  1.62313574547 	 & 	 $\times 10^{-10}$ 	  
& 	  2.40079049220 	 & 	 $\times 10^{-8}$ 	  
& 	  1.51029853345 	 & 	 $\times 10^{-9}$
& 	 -51 	 & 	 93 \\
 	 & 	 3 	 
& 	  2.44084848389 	 & 	 $\times 10^{-7}$ 	  
& 	  6.50157912447 	 & 	 $\times 10^{-10}$ 	  
& 	  2.91633622588 	 & 	 $\times 10^{-6}$ 	  
& 	  6.23354901544 	 & 	 $\times 10^{-9}$
& 	 -34 	 & 	 106 \\
 	 & 	 4 	 
& 	  1.12530626433 	 & 	 $\times 10^{-5}$ 	  
& 	  4.66621235553 	 & 	 $\times 10^{-9}$ 	  
& 	  1.37037638198 	 & 	 $\times 10^{-4}$ 	  
& 	  4.59633939401 	 & 	 $\times 10^{-8}$
& 	 -31 	 & 	 114 \\
\hline
5 	 & 	 0 	 
& 	  2.93546198223 	 & 	 $\times 10^{-15}$ 	  
& 	  1.68762144246 	 & 	 $\times 10^{-14}$ 	  
&    \multicolumn{2}{c|}{0}
&    \multicolumn{2}{c|}{0}
& 	 -94 	 & 	 94 \\
 	 & 	 1 	 
& 	  1.66341467681 	 & 	 $\times 10^{-13}$ 	  
& 	  2.73842758121 	 & 	 $\times 10^{-13}$ 	  
& 	  1.99357707469 	 & 	 $\times 10^{-12}$ 	  
& 	  2.10992243393 	 & 	 $\times 10^{-12}$
& 	 -77 	 & 	 92 \\
 	 & 	 2 	 
& 	  1.72172010497 	 & 	 $\times 10^{-12}$ 	  
& 	  1.49178217605 	 & 	 $\times 10^{-12}$ 	  
& 	  2.59625132235 	 & 	 $\times 10^{-11}$ 	  
& 	  1.34307539131 	 & 	 $\times 10^{-11}$
& 	 -63 	 & 	 100 \\
 	 & 	 3 	 
& 	  1.73935003471 	 & 	 $\times 10^{-9}$ 	  
& 	  1.01021973779 	 & 	 $\times 10^{-11}$ 	  
& 	  2.26258058740 	 & 	 $\times 10^{-8}$ 	  
& 	  9.56603438989 	 & 	 $\times 10^{-11}$
& 	 -46 	 & 	 109 \\
 	 & 	 4 	 
& 	  9.01787571564 	 & 	 $\times 10^{-8}$ 	  
& 	  3.64807139949 	 & 	 $\times 10^{-11}$ 	  
& 	  1.06079902733 	 & 	 $\times 10^{-6}$ 	  
& 	  3.50623060712 	 & 	 $\times 10^{-10}$
& 	 -29 	 & 	 121 \\
 	 & 	 5 	 
& 	  3.74854353561 	 & 	 $\times 10^{-6}$ 	  
& 	  3.02291684853 	 & 	 $\times 10^{-10}$ 	  
& 	  4.47051998131 	 & 	 $\times 10^{-5}$ 	  
& 	  2.96568439531 	 & 	 $\times 10^{-9}$
& 	 -19 	 & 	 130 \\
\hline
\hline
\multicolumn{2}{c||}{Total} 
& 	 2.10242675876 & $ \times 10^{-4} $ 	 
& 	 2.27174892328 & $ \times 10^{-6} $ 	 
& 	 2.75262625234 & $ \times 10^{-3} $ 	 
& 	 2.22779475534 & $ \times 10^{-5} $ \\
\hline
\hline
\end{tabular}
\caption{Energy and angular momentum fluxes for an eccentric orbit 
with $p = 8.75455$, $e = 0.764124$.  Note that we have folded the negative 
$m$ modes onto the corresponding positive $m$ modes and doubled the flux 
values in this table for $m > 0$.
\label{highEFluxTable}
}
\end{table*}

As expected, our code is more efficient for low eccentricities.  The first orbit
in Table \ref{totalFluxTable} ($p = 7.50478$, $e = 0.188917$), runs
in under a half hour on a single processor machine, 
giving the total flux for all 
$2 \le \ell \le 23$ (although note the limits on $m$ 
and $n$ mentioned in the previous subsection) to a
fractional error of $10^{-12}$.  
As $e$ increase, though, run times increase greatly.  The
second orbit in that table ($p = 8.75455$, $e = 0.764124$) takes 
six hours to achieve the same accuracy for all necessary $2 \le \ell \le 29$.
And, when $e=0.9$ for $2 \le \ell \le 20$ 
in the last row of Table \ref{Fujita2}, we had to 
raise our fractional error to $10^{-10}$ in order to get a run time of
 eighteen hours.  

Clearly, as $e$ gets close to 1, FD methods will lose out to TD codes, which
handle high eccentricities with more ease.  Still for $0 \le e \lesssim 0.9$,
our run times are not unreasonable when considering the high accuracy we 
achieve.

\section{Reconstruction of the metric perturbation amplitudes}
\label{reconstruct}

The full benefit of having complete and highly converged solutions for the 
master functions lies in using them to reconstruct the metric.  Ultimately, 
one wants to use the information, along with an appropriate regularization 
scheme, to compute the self force.  A developed approach to doing this is 
the mode-sum regularization method~\cite{Barack_1999}, which makes use of 
Lorenz gauge.  Here we use the information encoded in the master functions
to compute accurately the spherical harmonic amplitudes of the metric 
perturbation in Regge-Wheeler gauge.  The ability to determine the metric 
at all locations, including at the particle location, should serve as a 
useful starting point for computing the SF, either via a gauge transformation 
or an alternative regularization technique.  

We summarize the metric perturbation (MP) formalism in App.~\ref{MP}, where 
the definitions of the master functions, $\Psi^{\rm even}_{\ell m}$ and 
$\Psi^{\rm odd}_{\ell m}$, are given in terms of spherical harmonic amplitudes 
of the metric and their radial derivatives.  We reserve for this section 
giving the equations, (\ref{eq:ReconstructEven}) and 
(\ref{eq:ReconstructOdd}), for reconstructing the metric amplitudes in 
Regge-Wheeler gauge from the master functions.  These equations involve 
first derivatives, and in some cases second derivatives, of the master 
functions.  They also involve spherical harmonic projections of the 
stress-energy tensor.  Based on the form (\ref{eq:weakPsi}) anticipated 
in a master function, both of the abovementioned facts contribute to an 
expectation that the MP amplitudes might have point-singular behavior at 
$r_p(t)$ in the form of both $\d$ and $\d'$ terms.  We show that all 
potential $\d'$ terms cancel out.  However, in general a MP amplitude might 
have a functional form
\be
M(t,r) = M^{+}(t,r) \, \th(z) + M^{-}(t,r) \, \th(-z) + M^{S}(t) \, \d (z),
\q \q z \equiv r - r_{p}(t),
\label{eq:generalMP}
\ee
where $M^{+}$ ($M^{-}$) represents a smooth function in the region 
$r > r_{p}$ ($r < r_{p}$), and $M^S$ is a smooth function of $t$ alone,
giving the magnitude of the singularity.  We examine $M^{S}$ for all six 
non-zero MP amplitudes in the Regge-Wheeler gauge, and find three such 
terms to be nonvanishing.  Throughout the rest of this section we again 
suppress spherical harmonic labels $\ell$ and $m$.

As mentioned the metric reconstruction equations, of each parity, require 
spherical harmonic projections of the stress-energy tensor.  For a particle 
of mass $\mu$, traveling on a geodesic of the background spacetime, with 
four-velocity $u^\mu$, it is
\be
T^{\mu \nu} \l x^\a \r 
= \mu \int   \frac{d \tau}{\sqrt{-g}} 
u^\mu (\tau) u^\nu (\tau) \, \d^4 \left[ x - x_p(\tau) \right] .
\ee
In Schwarzschild coordinates the determinant of the metric is 
$g = -r^4 \sin^2 \th$.  After changing the variable of integration 
to coordinate time $t$, we have
\be
T^{\mu \nu} \l x^\a \r 
= \frac{\mu \, u^\mu (t) u^\nu (t) }{u^{t}(t) \ r_p(t)^2 }
\, \d [r - r_p(t)] 
\, \d [\varphi - \varphi_p(t)] 
\, \d [\th - \pi/2] .
\label{eq:SET}
\ee
Spherical harmonic projections of $T^{\mu \nu}$ appear as source terms 
in the decomposed Einstein equations (App.~\ref{MP}) and these are in turn
combined to produce the source terms for the master equations 
(App.~\ref{GF_Evaluated}).  In the subsections that follow, we evaluate 
the time dependence of all of the stress-energy tensor projections.
We use the definitions
\be
\La(r) \equiv \la + \frac{3M}{r},
\q \q
\la \equiv \frac{ \l \ell+2 \r \l \ell-1 \r }{2}. 
\ee 

\subsection{Even parity}
The even parity MP amplitudes are expressed in terms of $\Psi_{\rm even}$ 
and the source terms by (see \cite{Martel_2004})
\begin{align}
\label{eq:ReconstructEven}
\begin{split}
K (t,r) & =  f \pa_r \Psi_{\rm even} 
+ A \, \Psi_{\rm even} 
- \frac{r^2 f^2}{(\la + 1) \La} Q^{tt}, \\
h_{rr} (t,r) &= \frac{\La}{f^2}
\left[ \frac{\la+1}{r} \Psi_{\rm even} 
- K \right] + \frac{r}{f} \pa_r K, \\
h_{tr} (t,r) &=   r \pa_t \pa_r \Psi_{\rm even} 
+ r B \, \pa_t \Psi_{\rm even}  - \frac{r^2}{\la + 1} \left[ Q^{tr} 
+ \frac{rf}{\La} \pa_t Q^{tt} \right], \\
h_{tt} (t,r) &= f^2 h_{rr} +  f Q^\sharp,
\end{split}
\end{align}
where
\begin{align}
A(r) \equiv \frac{1}{r \La} 
\left[ \la(\la+1) + \frac{3M}{r} \l \la + \frac{2M}{r} \r \right], 
\q \q
B(r) \equiv \frac{1}{r f \La} 
\left[ \la \l 1 - \frac{3M}{r} \r - \frac{3M^2}{r^2}  \right].
\end{align}
These equations result from the definition (\ref{eq:masterEven}) of 
$\Psi_{\rm even}$ and its substitution into the even-parity field equations 
(\ref{eq:evenfieldeqns}).  The even-parity projections of the 
stress-energy tensor that appear in the equations above are defined by 
Eqs.~(\ref{eq:Qs}).  By enforcing the delta function constraints, they can 
be written in fully evaluated form (see App.~\ref{GF_Evaluated}), with each 
having a time dependent magnitude multiplying the radial delta function
\begin{align}
\begin{split}
Q^{ab} (t,r) &\equiv q^{ab} (t) \, \d [r - r_p(t)],
\q \q
Q^a (t,r) \equiv  q^{a} (t) \, \d [r - r_p(t)], \\
Q^\flat (t,r) &\equiv q^{\flat} (t) \, \d [r - r_p(t)],
\q \q
 Q^\sharp (t,r) \equiv q^{\sharp} (t) \, \d [r -r_p(t)] ,
\end{split}
\end{align}
where we use a lowercase $q$ as the base symbol of the corresponding 
magnitude.  With Eq.~(\ref{eq:fourVelocity}) giving the four-velocity 
$u^{\mu}$, the stress-energy tensor and Eqs.~(\ref{eq:Qs}) can be used to
find
\begin{align}
\begin{split}
\label{eq:qs}
q^{tt} (t) &= 8 \pi \mu \frac{\cal E}{r_p^2 f_{p}}  Y^{*}, 
\q \q 
q^{rr} (t) = 8 \pi \mu \frac{f_{p}}{{\cal{E}} r_p^2} 
\l {{\cal{E}}}^{2} - U_{p}^2  \r Y^{*}, 
\q \q
q^{tr} (t) =  8 \pi \mu \frac{u^{r}}{r_p^2}   Y^{*}, \\
q^{t}(t) &= \frac{16 \pi \mu}{\ell(\ell+1)} \frac{{\cal L}}{r_p^2} 
 Y_\varphi^{*}, 
\q \q
 q^{r}(t) = \frac{16 \pi \mu}{\ell(\ell+1)} 
\frac{{\cal{L}}}{{\cal{E}}} \frac{f_{p}}{r_p^2} u^r  Y_\varphi^{*}, \\
q^{\flat} (t) &= 8 \pi \mu \frac{{\cal{L}}^2}{{\cal{E}}} 
 \frac{f_{p}}{r_p^4}  Y^{*},
\q \q q^{\sharp} (t) = 32 \pi \mu \frac{(\ell-2)!}{(\ell+2)!} 
\frac{{\cal{L}}^2}{{\cal{E}}} \frac{f_{p}}{r_p^2}  
 Y^{*}_{\varphi \varphi}. 
\end{split}
\end{align}
Here, $Y$, $Y_{\varphi}$, and $Y_{\varphi \varphi}$ are shorthand for the 
even-parity scalar, vector, and tensor spherical harmonics, respectively, 
evaluated along the worldline at $\th = \pi/2$ and $\varphi = \vp_{p}(t)$.

Now consider the reconstruction of the MP amplitude $K$, given
in Eq.~(\ref{eq:ReconstructEven}).  Using the expected functional form of 
$\Psi$ given in Eq.~(\ref{eq:weakPsi}), $K$ obviously does fit the general
form (\ref{eq:generalMP}) claimed above.  In fact, we find
\begin{align}
K^{\pm} (t,r) =  f \pa_r \Psi^{\pm} + A \Psi^{\pm},
\q \q
K^{S} (t) = f_{p} \llbracket \Psi \rrbracket_{p} 
- \frac{r_{p}^2 f_{p}^2}{(\la + 1) \La_{p}} q^{tt} = 0 ,
\end{align}
where the vanishing of $K^{S}$ follows from use of Eq.~(\ref{eq:psiJump}) 
for $\llbracket \Psi \rrbracket_{p}$, and $q^{tt}$ from Eq.~(\ref{eq:qs}).  
Therefore, we see that the even-parity metric function $K$ in Regge-Wheeler 
gauge is (only) a $C^{-1}$ function at the location of the particle.

Using the same approach to evaluate $h_{rr}$ in Eq.~(\ref{eq:ReconstructEven})
we have
\begin{align}
h_{rr}^{\pm} (t,r) =  \frac{\La}{f^2} \left[ \frac{\la+1}{r} \Psi^{\pm}
- K^{\pm} \right] + \frac{r}{f} \pa_r K^{\pm}, 
\q \q
h_{rr}^{S} (t) = \frac{r_{p}}{f_{p}} \llbracket K \rrbracket_{p}
=  r_{p} \llbracket \pa_r \Psi \rrbracket_{p} 
 + \frac{r_{p} A_{p}}{f_{p}} \llbracket \Psi \rrbracket_{p} .
 \label{eq:hrrS}
\end{align}
Here, we have extended in a natural way the use of the 
$\llbracket \ \ \rrbracket_{p}$ notation to let 
$\llbracket K \rrbracket_{p}$ represent the jump in $K$ at $z=0$.  We 
find that the Regge-Wheeler metric function $h_{rr}$ is not only 
discontinuous across $r_p(t)$ but also has a point-singular term, which is 
an artifact of Regge-Wheeler gauge.

The $h_{tr}$ function is more subtle than the previous two.  Looking at 
Eq.~(\ref{eq:ReconstructEven}), we need the following terms involving 
$\Psi$, 
\begin{align}
\begin{split}
r B \, \pa_t \Psi &=  r B \, \pa_t \Psi^+ \, \th (z) 
+ r B \, \pa_t \Psi^- \, \th (-z) 
- r_{p} B_{p} \dot r_p \llbracket \Psi \rrbracket_{p} \, \d (z), \\
r \pa_{t} \pa_{r} \Psi &= 
r \pa_{t} \pa_{r} \Psi^+ \th (z) + 
r \pa_{t} \pa_{r} \Psi^- \th (-z) 
+ \left[ r_{p} \frac{d}{dt} \Big( \llbracket \Psi \rrbracket_{p} \Big)
+ \dot r_p \llbracket \Psi \rrbracket_{p}
- r_{p} \dot r_p  \llbracket \pa_r \Psi \rrbracket_{p} \right] \d (z) 
- r_{p} \dot r_p \llbracket \Psi \rrbracket_{p} \, \d' (z).
\end{split}
\end{align}
On the right side of these equations we have evaluated all the $\d$ and 
$\d'$ coefficients at $z=0$ with Eqs.~(\ref{eq:deltaEvalR}) and 
(\ref{eq:deltaPrimeEvalR}) (fully evaluated form).  The singular terms that 
arise in these expressions can be grouped with the similarly singular
contributions from the source terms,
\begin{align}
\begin{split}
\frac{r^{2}}{\la + 1} Q^{tr}
&= \frac{r_{p}^{2}}{\la + 1} q^{tr} \d(z), \\
\frac{r^{3} f}{(\la + 1) \La} \pa_{t} Q^{tt} 
&= \frac{1}{(\la + 1) \La_{p}}
\left[
r_{p}^{3} f_{p} \frac{d q^{tt}}{dt}
+ \frac{3 \la r_{p}^{2} + 12 M r_{p} - 4\la  M r_{p}-18 M^{2}}{\La_{p}}
\dot r_{p} q^{tt}
\right] \d(z)
- \frac{r_{p}^{3} f_{p}}{(\la + 1) \La_{p}} \dot r_{p} q^{tt} \, \d'(z).
\end{split}
\end{align}
Upon carefully checking the time dependence of $q^{tt}$ and the jump in 
$\Psi$, we find that the $\d'$ terms cancel out.  There are multiple $\d$ 
terms, but after using the expressions for $\llbracket \Psi \rrbracket_{p}$, 
$\llbracket \pa_{r}\Psi \rrbracket_{p}$ in (\ref{eq:psiJump}) and the 
relevant $q$'s in (\ref{eq:qs}), most of the terms cancel and we are left 
with
\begin{align}
h_{tr}^{\pm} (t,r) = r \pa_{t} \pa_{r} \Psi^{\pm} + r B \, \pa_{t} \Psi^{\pm},
\q \q
h_{tr}^{S} (t) = {\cal{E}}^2 \frac{\dot r_p}{ f_{p} U_{p}^2} q^{\sharp} .
\label{eq:htrS}
\end{align}

Finally, the $h_{tt}$ term is simple.  We insert Eq.~(\ref{eq:hrrS})
into the field equation for $h_{tt}$ and get
\begin{align}
h_{tt}^{\pm} (t,r) = f^{2} h_{rr}^{\pm},
\q \q
h_{tt}^{S} (t) = f_{p}^{2} h_{rr}^{S}
+  f_{p} q^\sharp.
\label{eq:httS} 
\end{align}
So, we see that in Regge-Wheeler gauge $K$ is $C^{-1}$ with no singularity 
along the worldline of the particle, but the three even-parity MP amplitudes
in the ``$t,r$ sector'' have point-singular artifacts given by 
Eqs.~(\ref{eq:hrrS}), (\ref{eq:htrS}), (\ref{eq:httS}).

\subsection{Odd parity}
Once $\Psi_{\rm odd}$ has been computed, the odd-parity MP amplitudes 
can be reconstructed via 
\begin{align}
\label{eq:ReconstructOdd}
h_t (t,r) = \frac{f}{2} \pa_r \l r \Psi_{\rm odd} \r 
- \frac{r^2 f}{2 \la } P^t,
\q \q
h_r (t,r) = \frac{r}{2 f} \pa_t \Psi_{\rm odd} 
+ \frac{r^2}{2 \la f} P^r ,
\end{align}
(see \cite{Lousto_2005}).  These equations follow from the definition 
(\ref{eq:masterOdd}) and its substitution into the odd-parity field 
equations (\ref{eq:oddfieldeqns}).  Similar to before, we define the 
lowercase $p$'s to be the time-dependent magnitudes of the radial delta 
function after fully evaluating the odd-parity projections of the 
stress-energy tensor
\begin{align}
P^{a} (t,r) \equiv p^{a} (t) \, \d [r - r_{p}(t)] ,
\q \q
 P  (t,r)  \equiv p (t) \, \d [r - r_{p}(t)].
\end{align}
Also as before, we use the time dependence of the four-velocity and the 
stress-energy tensor to determine these magnitudes for eccentric motion on 
Schwarzschild, 
\begin{align}
\label{eq:ps}
p^{t} (t) = \frac{16 \pi \mu}{\ell (\ell + 1)} 
\frac{{\cal{L}}}{r_{p}^{2}} 
 X^{*}_{\varphi}, 
\q \q
p^{r} (t) = \frac{16 \pi \mu}{\ell (\ell + 1)} 
\frac{{\cal{L}}}{{\cal{E}}}
\frac{f_{p}}{r_{p}^{2}} u^{r}  X^{*}_{\varphi} ,
\q \q
p (t) = 16 \pi \mu \frac{(\ell - 2)!}{(\ell + 2)!} 
\frac{{\cal{L}}^{2}}{{\cal{E}}} \frac{f_{p}}{r_{p}^{2}} 
 X^{*}_{\varphi \varphi}.
\end{align}
Here, $X_{\varphi}$ and $X_{\varphi \varphi}$ are shorthand for the 
odd-parity vector and tensor spherical harmonics evaluated along the 
worldline at $\th = \pi/2$ and $\varphi = \vp_{p}(t)$.

Now, as in the even-parity case we can analyze the local structure of the 
MP amplitudes.  We again assume $\Psi$ to have the form 
Eq.~(\ref{eq:weakPsi}).  Plugging the relevant expressions into 
Eq.~(\ref{eq:ReconstructOdd}) for the odd-parity MP amplitude reconstruction,
we find that all the point-singular parts cancel out exactly, leaving
\begin{align}
\begin{split}
h_t^{\pm} (t,r) = \frac{f}{2} \pa_r \l r \Psi^{\pm} \r, 
\q \q
h^{\rm S}_t (t) = 0, \\
h_r^{\pm} (t,r) = \frac{r}{2 f} \pa_t \Psi^{\pm}, 
\q \q 
h^{\rm S}_r (t) = 0.
\end{split}
\end{align}
So, we see that the odd-parity MP functions in Regge-Wheeler gauge are 
smooth as they approach $r_p(t)$ with only a finite jump at that point.  

\begin{figure}[h!]
{
	\begin{center}
	{
		\includegraphics{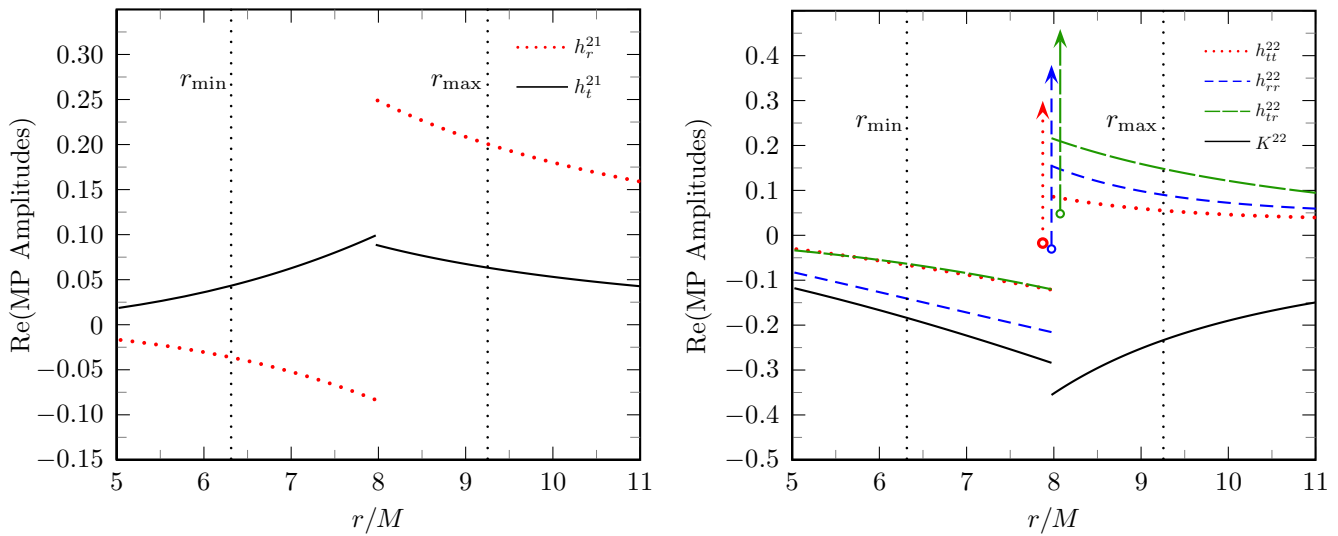}
		\caption
		{ 
			\label{fig:MPGraph}
			The EHS approach to
			reconstructing the TD MP amplitudes.
			We consider a particle orbiting
			with $p = 7.50478$ and $e = 0.188917$  at $t = 80.62 M$.
			The left plot shows the odd-parity MP amplitudes $h^{21}_{r}$ and 
			$h^{21}_{t}$.  The right shows the even-parity 
			$h^{22}_{tt}$, $h^{22}_{rr}$, $h^{22}_{tr}$, and  $K^{22}$. 
			Note that the amplitudes $h^{22}_{tt}$, $h^{22}_{rr}$, 
			and $h^{22}_{tr}$ are 
			singular along the particle's worldline, as indicated by
			arrows in the plot on the right.  The magnitude
			of these singularities are given in Eqs.~(\ref{eq:hrrS}), 
			(\ref{eq:htrS}), (\ref{eq:httS}).
			The remaining three MP amplitudes approach the particle 
			location smoothly, and have only a finite jump at 
			that point.
		}

	}
	\end{center}
}
\end{figure}

FIG.~\ref{fig:MPGraph} summarizes these findings graphically, for both even 
and odd parity, using several specific spherical harmonic modes.

\section{Conclusion}

We have achieved two main results with this paper.  First, we have shown 
successful application of the method of extended homogeneous solutions to 
gravitational perturbations from a small mass in eccentric orbit about a 
massive Schwarzschild black hole.  In doing so, we accurately computed the 
master functions in the Regge-Wheeler-Zerilli formalism in the frequency 
domain and transformed these fields back to the time domain.  With this 
method we achieved exponential convergence of the master functions and 
their derivatives for all $r$ including the instantaneous particle location 
$r=r_p(t)$.

Our second important result is the reconstruction of the metric perturbation 
amplitudes in Regge-Wheeler gauge for arbitrary radiative modes.  In addition 
to computing the smooth parts of these amplitudes, we have derived the 
time dependent magnitudes of point-singular terms that reside at $r_p(t)$ 
in some components of the metric.  This full and accurate knowledge of the 
spherical harmonic amplitudes of the metric at, and near, $r_p(t)$ lays the 
groundwork for one or more subsequent approaches to local computation of the 
self-force.

\acknowledgments

We thank Steven Detweiler, Scott Hughes, Paul Anderson, and Stephen Lau 
for helpful discussions.  We thank the referee also for several valuable 
suggestions.  SH acknowledges support from the US Department 
of Education GAANN fellowship number P200A090135 and the NC Space Grant's 
Graduate Research Assistantship Program.  CRE acknowledges support from 
the Bahnson Fund at the University of North Carolina--Chapel Hill.

\appendix

\section{The fully evaluated form of distributional source terms}
\label{generalized}

In the RWZ formalism for perturbations generated by an orbiting point mass, 
the master equations have distributional sources with both delta function 
and derivative-of-delta function terms.  Reduced by spherical harmonic 
decomposition, these distributions have support only along a 
one-dimensional timelike worldline $r=r_p(t)$ within a two dimensional 
domain.  The delta function's behavior is still elementary, 
\be
\label{eq:deltaEvalR}
\a(t,r) \, \d [r - r_p(t)]  
= \a \left(t, r_p(t) \right) \, \d [r - r_p(t)] 
\equiv
\tilde \a (t)  \, \d [r - r_p(t)]  ,
\ee
where $\a(t,r)$ is assumed to be a smooth function and we use the 
notation $\tilde \a (t)$ to indicate the one-dimensional function that 
results from restricting (or fully evaluating) $\a (t,r)$ with the delta 
function.  At any stage in a calculation a delta function can be used to
fully evaluate all smooth functions that multiply it.  Under an integral 
the result is obvious
\be
\int \a(t,r) \, \d [r - r_p(t)] \, dr = \tilde \a (t) ,
\ee
with the resulting function of time being unique.  Occasionally, there is 
need to differentiate such a function.  The total derivative is related 
to derivatives of the original function by
\be
\frac{d\tilde\a}{dt} = \Big[ \pa_t \a(t,r) + {\dot r}_p \pa_r \a(t,r) 
\Big]_{r=r_p(t)} ,
\ee
where on the right hand side we differentiate first and evaluate second.

Of more interest is the behavior of $\d'$ \cite{Lighthill_1958}.  
Differentiating Eq.~(\ref{eq:deltaEvalR}) with respect to $r$, we obtain
\be
\a(t, r) \, \d' [r - r_p(t)] + \pa_r \a(t,r) \, \d [r-r_p(t)]
= \tilde \a (t) \,  \d' [r - r_p(t)] .
\ee
Rearranging terms and using the rule of fully evaluating whenever possible, 
we find 
\be
\label{eq:deltaPrimeEvalR}
\a(t, r) \, \d' [r - r_p(t)] 
= \tilde \a (t) \,  \d' [r - r_p(t)] 
- \tilde \b(t) \, \d [r- r_p(t)], 
\q \q \tilde \b(t) 
\equiv \pa_r \a (t,r_p(t)) 
\equiv \Big[ \pa_r \a (t, r) \Big]_{r = r_p(t)} ,
\ee
which is the analogous fully evaluated form.  Upon integration,
\be
\int \a(t, r) \, \d' [r - r_p(t)] \, dr = 
- \tilde \b(t) =
- \pa_r \a (t, r_p(t)) .
\ee
Since the first term on the right of Eq.~(\ref{eq:deltaPrimeEvalR}) 
disappears upon integration, why retain it?  The answer is that we may 
multiply Eq.~(\ref{eq:deltaPrimeEvalR}) by another smooth (test) function,
$\g(t,r)$.  We can then proceed to fully evaluated form by reducing the 
smooth function $\g(t,r)\, \a(t,r)$ on the left or use the same reduction 
on the first term on the right.  In either case the result is
\be
\label{eq:deltaPrimeEvalR2}
\g(t,r)\, \a(t, r) \, \d' [r - r_p(t)] 
= \tilde \g(t)\, \tilde \a (t) \, \d' [r - r_p(t)] 
- \tilde \a(t)\, \pa_r \g(t,r_p(t)) \, \d [r- r_p(t)]
- \tilde \g(t)\, \pa_r \a(t,r_p(t)) \, \d [r- r_p(t)] .
\ee
From this it is evident that we can \emph{partially} evaluate a coefficient 
of $\d'$ in a number of different ways.  

Martel \cite{Martel_2004} introduced the notation found in 
Eq.~(\ref{eq:SMartelForm}) for gravitational master function source terms, 
with two-dimensional functions $G_{\ell m} (t,r)$ and $F_{\ell m} (t,r)$ 
multiplying $\d$ and $\d'$, respectively.  In examining the Zerilli-Moncrief 
master function, he left these coefficients partially evaluated.  Sopuerta 
and Laguna \cite{SL_2008} started with the same notation for 
$G_{\ell m}(t,r)$ and $F_{\ell m} (t,r)$ in the case of the 
Cunningham-Price-Moncrief master function, and fully evaluated these 
coefficients at $r=r_p(t)$.  A difficulty with the $G_{\ell m} (t,r)$ and 
$F_{\ell m} (t,r)$ notation is that there is no unique form of these 
functions if partially evaluated.  Any solution of the RWZ wave equation 
will require a full evaluation of the source.  The procedure should not 
matter but we prefer the clarity afforded by using the identities found 
in Eqs.~(\ref{eq:deltaEvalR}) and (\ref{eq:deltaPrimeEvalR}) to write 
Eq.~(\ref{eq:SMartelForm}) in fully evaluated form from the outset 
\be
S_{\ell m}(t,r) = 
\tilde G_{\ell m} (t) \ \d [r - r_p(t)] + 
\tilde F_{\ell m} (t) \ \d' [r - r_p(t)] ,
\ee
where
\begin{align}
\label{eq:GFtilde}
\tilde G_{\ell m} (t) 
\equiv \Big[ 
G_{\ell m} (t , r) - \pa_r F_{\ell m} (t , r) \Big]_{r = r_p(t)}, 
\q \q
\tilde F_{\ell m} (t) 
 \equiv \Big[ F_{\ell m} (t , r) \Big]_{r = r_p(t)}.
\end{align}

\section{Source terms for eccentric motion on Schwarzschild}
\label{GF_Evaluated}

Here we give the unambiguous expressions for 
$\tilde G_{\ell m}$ and $\tilde F_{\ell m}$ 
for the even-parity Zerilli-Moncrief and odd-parity 
Cunningham-Price-Moncrief master functions fully 
evaluated at $r = r_p(t)$.  We introduce new notation for constituent 
parts of $\tilde G_{\ell m}$ and $\tilde F_{\ell m}$ based upon the
projections of the stress-energy tensor defined in App.~\ref{MP} and the
fully evaluated time-dependent magnitudes of $\d [r-r_p(t)]$ given by 
Eqs.~(\ref{eq:qs}) and (\ref{eq:ps}).  Note that we use ${\cal G}$ and 
${\cal F}$ to denote additional time-dependent factors that multiply the 
various stress-energy magnitudes.  The indices on these ${\cal G}$ and 
${\cal F}$ factors are not tensor indices.

\subsection{Even parity}
In the even-parity case, we examine the terms first published by Martel 
\cite{Martel_2004}, but now fully evaluate them at $r=r_{p}(t)$.
We find,

\begin{align}
\begin{split}
\tilde G_{\ell m} (t)
&= {\cal{G}}_{\ell}^{rr} \, q_{\ell m}^{rr}  
+ {\cal{G}}_{\ell}^{tt} \, q_{\ell m}^{tt} 
+ {\cal{G}}_{\ell}^{r} \, q_{\ell m}^{r} 
 + {\cal{G}}_{\ell}^{\flat} \, q_{\ell m}^{\flat}
  + {\cal{G}}_{\ell}^{\sharp} \, q_{\ell m}^{\sharp}  \\
\tilde F_{\ell m} (t)
&= {\cal F}_{\ell}^{rr} \, q_{\ell m}^{rr}
 + {\cal F}_{\ell}^{tt} \, q_{\ell m}^{tt} ,
\end{split}
\end{align}
where 
\begin{align}
\begin{split}
{\cal G}_{\ell}^{rr} (t) & \equiv \frac{1}{\l \la + 1 \r r_p \La_{p}^{2}}
\Big[ \l \la + 1 \r \l \la r_p + 6M \r r_p + 3M^{2} \Big], \\
{\cal G}_{\ell}^{tt} (t) & \equiv - \frac{f_{p}^{2}}{(\la + 1) r_p \La_{p}^{2}}
 \Big[ \la \l  \la + 1 \r r_p^2 + 6 \la Mr_p   + 15 M^2 \Big], \\
{\cal G}_{\ell}^{r} (t) & \equiv \frac{2f_{p}}{\La_{p}}, 
\q \q
{\cal G}_{\ell}^{\flat} (t) \equiv \frac{r_p f_{p}^{2}}{ (\la+1) \La_{p}},
\q \q
{\cal G}_{\ell}^{\sharp} (t) \equiv - \frac{f_{p}}{r_p}, \\
{\cal F}_{\ell}^{rr} (t) & \equiv - \frac{r_p^{2} f_{p}}{\l \la + 1 \r \La_{p}}, 
\q \q 
{\cal F}_{\ell}^{tt} (t) \equiv \frac{r_p^{2} f_{p}^{3}}{\l \la + 1 \r \La_{p}},
\end{split}
\end{align}
with the $q$'s given in Eq.~(\ref{eq:qs}).

\subsection{Odd parity}
In the odd-parity case, the fully evaluated source magnitudes are equivalent 
to those first published by Sopuerta and Laguna \cite{SL_2008} and later 
with more detail by Field, Hesthaven, and Lau \cite{FHL_2009}.  We find,
\begin{align}
 \tilde G_{\ell m} (t) = 
{\cal{G}}_{\ell}^{r_{1}}  \, p^{r}_{\ell m}  
+ {\cal{G}}_{\ell}^{r_{2}}  \, \frac{d p^{r}_{\ell m}}{dt}   
+ {\cal{G}}^{t}_{\ell}  \, p_{\ell m}^{t},  
\q \q
\tilde F_{\ell m} (t) = 
{\cal{F}}_{\ell}^{r}  \, p_{\ell m}^{r} 
+ {\cal{F}}_{\ell}^{t}  \, p_{\ell m}^{t},
\end{align}
where
\begin{align}
{\cal{G}}_{\ell}^{r_{1}} (t)  \equiv \frac{\dot r_{p}}{\la},
\q \q
{\cal{G}}_{\ell}^{r_{2}} (t)  \equiv \frac{r_{p}}{\la},
\q \q
 {\cal{G}}_{\ell}^{t} (t)  \equiv - \frac{f_{p}}{\la},
\q \q
 {\cal{F}}_{\ell}^{r} (t)  \equiv - \frac{r_{p} \dot r_{p}}{\la},
\q \q
 {\cal{F}}_{\ell}^{t} (t)  \equiv \frac{r_{p} f_{p}^{2}}{\la} ,
\end{align}
and the $p$'s are given by Eq.~(\ref{eq:ps}).

\section{Metric perturbation formalism in the Regge-Wheeler gauge}

\label{MP}

Here we briefly summarize the definitions of metric perturbation 
(MP) amplitudes (on a common tensor spherical harmonic basis) for both even 
and odd parities.  The field equations and Bianchi identities are given in 
terms of the MP amplitudes and spherical harmonic projected source terms.  
The specific gauge invariant master functions we use in our simulations 
are expressed in terms of the MP amplitudes and their associated master 
equations, potentials, and source terms are summarized.  In what follows, 
lowercase Latin indices will run over $(t, r)$, while uppercase Latin 
indices will run over $(\th, \varphi)$.  
This section draws heavily from Martel and Poisson \cite{MP_2005}.  The 
material here serves as a basis for discussing in Sec.~\ref{reconstruct}
how the MP can be numerically reconstructed from the master functions.

\subsection{Even parity}
Of the ten MP amplitudes, seven are in the even-parity sector.
Using the decomposition of Martel and Poisson \cite{MP_2005}, they are 
\begin{align}
\label{eq:MPEven}
p_{ab} \l x^{\mu} \r &= \sum_{\ell, m} h_{ab}^{\ell m} Y^{\ell m}, 
& p_{aB}  \l x^{\mu} \r &= \sum_{\ell, m} j_a^{\ell m} Y_B^{\ell m}, 
& p_{AB} \l x^{\mu} \r &= r^2 \sum_{\ell, m} \Big( K^{\ell m} \O_{AB} Y^{\ell m} 
+  G^{\ell m} Y^{\ell m}_{AB} \Big).
\end{align}
The tensor $\O_{AB}$ is the metric on the unit two-sphere,
\be 
 ds^{2} = \O_{AB} dx^{A} dx^{B} = d \th^{2} + \sin^{2} \th \, d \varphi^{2}.
 \ee
The even-parity scalar ($Y^{\ell m}$), vector ($Y^{\ell m}_{A}$), 
and tensor ($Y^{\ell m}_{AB}$ and $\O_{AB} Y^{\ell m}$) spherical
harmonics are defined in \cite{MP_2005}.  Note that $Y_{AB}^{\ell m}$ is 
the trace-free tensor spherical harmonic, which differs from what
Regge and Wheeler used in their original work \cite{RW_1957}.  
For the remainder of this section, we drop $\ell$ and $m$ 
indices for the sake of brevity.

In Schwarzschild coordinates, the amplitudes defined here are related to 
Regge and Wheeler's original quantities.  In the ``$t,r$ sector,'' 
$h_{tt} = f H_{0}$, 
$h_{tr} = H_{1}$,
and 
$h_{rr} = H_{2} / f$.  For the off-diagonal 
elements, $j_{t} = h_{0}$ and
$j_{r} = h_{1}$.  Finally, on the two-sphere 
$G_{\rm here} = G_{\rm RW}$, while
$K_{\rm here} = K_{\rm RW} 
- \ell (\ell + 1) G / 2$.  We use the Regge-Wheeler gauge, where 
$j_{a} = G = 0$.  In this gauge and in Schwarzschild coordinates, the 
even-parity field equations are
\begin{align}
\begin{split}
\label{eq:evenfieldeqns}
- \pa_r^2  K - \frac{3r - 5M}{r^2 f} \pa_r  K 
+ \frac{f}{r} \pa_r  h_{rr}
+\frac{\l \la + 2 \r r + 2M}{r^3}  h_{rr} + \frac{\la}{ r^2 f}  K 
&= Q^{tt}, \\
 \pa_t \pa_r  K 
+ \frac{r - 3M}{r^2 f} \pa_t  K - \frac{f}{r} \pa_t  h_{rr} 
- \frac{\la + 1}{ r^2}  h_{tr} &= Q^{tr}, \\
 - \pa_t^2  K 
+ \frac{(r-M)f}{r^2} \pa_r  K + \frac{2f}{r} \pa_t  h_{tr}
- \frac{f}{r} \pa_r  h_{tt} + \frac{(\la + 1)r + 2M}{r^3}  h_{tt} 
- \frac{f^2}{r^2}  h_{rr} 
- \frac{\la f}{r^2}  K &= Q^{rr}, \\
\pa_t  h_{rr} - \pa_r  h_{tr} 
+ \frac{1}{f} \pa_t  K - \frac{2M}{r^2 f}  h_{tr} &= Q^{t}, \\
-\pa_t  h_{tr} 
+ \pa_r  h_{tt} - f \pa_r  K - \frac{r - M}{r^2 f}  h_{tt} 
+ \frac{(r-M) f}{r^2}  h_{rr} &= Q^{r}, \\
\begin{split}
-\pa_t^2  h_{rr} 
+ 2 \pa_t \pa_r  h_{tr} - \pa_r^2  h_{tt} 
- \frac{1}{f} \pa_t^2  K + f \pa_r^2  K 
+ \frac{2 (r - M) }{r^2 f} \pa_t  h_{tr}
- \frac{r - 3M}{r^2 f} \pa_r  h_{tt} 
- \frac{(r-M) f}{r^2} \pa_r  h_{rr}  & \\
 + \frac{2(r-M)}{r^2} \pa_r  K 
+ \frac{( \la +1) r^2 - 2(\la + 2) M r + 2M^2}{r^4 f^2}  h_{tt}
- \frac{(\la+1) r^2 - 2\la M r - 2 M^2}{r^4}  h_{rr} &= Q^\flat,
\end{split} \\
\frac{1}{f}  h_{tt} - f  h_{rr} &= Q^\sharp ,
\end{split}
\end{align}
which rely upon the following source terms 
\begin{align}
\label{eq:Qs}
\begin{split}
Q^{ab} (t,r) &\equiv 8 \pi \int T^{ab} Y^{*} \, d \O,  
\q \q Q^a (t,r) \equiv \frac{16 \pi  r^2}{\ell (\ell +1)}
\int T^{aB} Y^{*}_B \, d \O, \\
Q^\flat (t,r)
&\equiv 8 \pi r^2 \int T^{AB} \O_{AB} Y^{*} \, d \O, 
\q \q Q^\sharp (t,r) 
\equiv 32 \pi r^4 \frac{(\ell - 2)!}{(\ell + 2)!}
\int T^{AB} Y^{*}_{AB} \, d \O. 
\end{split}
\end{align}
The conservation (Bianchi) identities are
\begin{align}
\begin{split}
\pa_t Q^{tt} + \pa_r Q^{tr} + 2 \frac{(r-M)}{r^2 f} Q^{tr} 
- \frac{\la + 1}{r^2} Q^t &= 0, \\
\pa_t Q^{tr} + \pa_r Q^{rr} + \frac{Mf}{r^2} Q^{tt} 
+ \frac{2r - 5M}{r^2 f} Q^{rr} - \frac{\la+1}{r^2} Q^r 
- \frac{f}{r} Q^\flat &= 0, \\
\pa_t Q^t + \pa_r Q^r + \frac{2}{r} Q^r + Q^\flat - \frac{\la}{r^2} 
Q^\sharp &= 0.
\end{split}
\end{align}
We use the gauge invariant \emph{Zerilli-Moncrief} master function
(see \cite{Moncrief_1974,CPM_1979}, modifying the approach of 
\cite{Zerilli_1970}), which is 
\be
\label{eq:masterEven}
\Psi_{\rm even} (t,r)
\equiv \frac{2r}{\ell (\ell +1)} \left[ K 
+ \frac{1}{ \La} 
\l f^2 h_{rr} - r f \pa_r K \r \right],
\ee
in Schwarzschild coordinates.
It satisfies the wave equation
\be
\left[ -\frac{\pa^2}{\pa t^2}  + \frac{\pa^2}{\pa r_*^2} 
- V_{\rm even} \right] \Psi_{\rm even} = S_{\rm even},
\ee
with source term 
\begin{multline}
S_{\rm even} (t,r) \equiv \frac{1}{\l \la + 1 \r \La} 
\bigg[  
r^2 f \l f^2 \pa_r Q^{tt} - \pa_r Q^{rr}\r 
+ r ( \La  - f ) Q^{rr} 
+ r f^2 Q^\flat \\
- \frac{f^2}{r \La} 
\Big( \la ( \la - 1 ) r^2 + \l 4 \la - 9 \r Mr + 15 M^2 \Big) Q^{tt} 
\bigg]
+ \frac{2 f}{\La} Q^r - \frac{f}{r} Q^\sharp 
,
\end{multline}
and standard Zerilli potential
\be
V_{\rm even} (r) \equiv 
\frac{f}{r^2 \La_{\rm}^2} \left[ 2 \la^2  \l \la+1 + \frac{3M}{r} \r 
+ \frac{18M^2}{r^2} \l \la + \frac{M}{r} \r \right] .
\label{eq:evenPotential}
\ee

\subsection{Odd parity}
The remaining three MP amplitudes belong to the odd-parity sector,
\begin{align}
\label{eq:MPOdd}
p_{ab} \l x^{\mu} \r = 0, 
\q \q 
p_{aB} \l x^{\mu} \r  = \sum_{\ell, m} h_a^{\ell m} X_B^{\ell m},  
\q \q
p_{AB} \l x^{\mu} \r  = \sum_{\ell, m} h_2^{\ell m} X^{\ell m}_{AB} .
\end{align}
The vector ($X_B^{\ell m}$) and tensor ($X_{AB}^{\ell m}$) spherical 
harmonics are those defined in \cite{MP_2005}.  Note that the tensor 
spherical harmonics differ from those used by Regge and Wheeler by a minus 
sign.  For the remainder of this section, we again drop $\ell$ and $m$ 
indices.

These MP amplitudes are related to Regge and Wheeler's quantities through
$ h_{t} = h_0 $,
$ h_{r} = h_1 $,
and 
$ h_{2}^{\rm here} = -h^{\rm RW}_2 $.
We use Regge-Wheeler gauge, in which $h_{2} = 0$.  In this gauge and in 
Schwarzschild coordinates, the odd-parity field equations are
\begin{align}
\begin{split}
\label{eq:oddfieldeqns}
 - \pa_t \pa_r  h_r + \pa_r^2  h_t - \frac{2}{r} \pa_t  h_r 
- \frac{2 (\la + 1) r - 4M}{r^3 f}  h_t &= P^t, \\
\pa_t^2  h_r - \pa_t \pa_r  h_t + \frac{2}{r} \pa_t  h_t 
+ \frac{2 \la f }{r^2}  h_r &= P^r, \\
 -\frac{1}{f} \pa_t  h_t + f \pa_r  h_r + \frac{2M}{r^2}  h_r &= P ,
\end{split}
\end{align}
with source terms given by
\begin{align}
P^a (t,r)
\equiv  \frac{16 \pi r^2}{\ell (\ell +1)}
\int T^{aB} X_{B}^{*} \, d\O ,
\q \q
P (t,r)  \equiv  
16 \pi r^4 \frac{(\ell - 2)!}{(\ell + 2)!} 
\int T^{AB} X^{*}_{AB}\, d\O.
\end{align}
The conservation (Bianchi) identity is
\be
\pa_t P^t + \pa_r P^r + \frac{2}{r} P^r - \frac{2 \la}{r^2} P = 0.
\ee
In the odd-parity sector, we use the gauge-invariant
\emph{Cunningham-Price-Moncrief} master function \cite{CPM_1978}, which in Schwarzschild coordinates is
\be
\Psi_{\rm odd}(t,r) \equiv \frac{r}{\la} 
\left[ \pa_r h_t  
- \pa_t  h_r - \frac{2}{r} h_{t} \right] .
\label{eq:masterOdd}
\ee
It satisfies the wave equation
\be
\left[ -\frac{\pa^2}{\pa t^2}  + \frac{\pa^2}{\pa r_*^2} 
- V_{\rm odd} \right] \Psi_{\rm odd} = S_{\rm odd},
\ee
with source term
\be
S_{\rm odd} (t,r) \equiv \frac{r f}{\la}  
\left[\frac{1}{f} \pa_t P^{r} + f \pa_r P^{t} + \frac{2M}{r^{2}} P^{t} \right] ,
\label{eq:SOdd} 
\ee
and standard Regge-Wheeler potential
\be
V_{\rm odd} (r) \equiv \frac{f}{r^2} 
\left[ \ell \l \ell + 1 \r - \frac{6M}{r} \right].
\label{eq:oddPotential}
\ee

\section{Asymptotic expansions for Jost functions at $r_* \to \infty$}

\label{asympExp}

We examine here the asymptotic expansions
that we use to set boundary conditions far from the black hole.  
The unit normalized solution to Eq.~(\ref{eq:FDwaveEq}) is factored
into the form
\be
\hat R^+_{\ell mn} (r) = J^+_{\ell mn} (r) e^{i \o_{mn} r_*} ,
\ee
where $J^+_{\ell mn}$ is the ``Jost function'' \cite{Chandra_1983}, which
goes to 1 as $r_* \to +\infty$.  (We can similarly define the horizon side 
Jost function  through 
$\hat R^-_{\ell mn} = J^-_{\ell mn} e^{-i \o_{mn} r_*}$,
which goes to 1 as $r_{*} \to - \infty$.)
Plugging this into the source free version of
Eq.~(\ref{eq:FDwaveEq}) and changing to $r$ 
derivatives, we have
\be
f \frac{d^{2} J^+_{\ell mn} }{d r^{2}}
+ \left[ \frac{ 2 M}{r^2} + 2 i \o_{mn} \right] \frac{d J^+_{\ell mn}}{dr}
- \frac{V_\ell }{f}  J^+_{\ell mn}  = 0 .
\ee
From here we assume an asymptotic series solution of 
$J^+_{\ell mn}$ of the form
\be
J^+_{\ell mn} (r) = \sum_{j = 0}^\infty \frac{a_j}{\l \o_{mn} r \r^j}
\ee
Note that contrary to a Taylor expansion which converges for 
fixed $r$ with increasing $j$, this series converges for fixed
$j$ with increasing $r$.
When a 
specific potential is chosen, the method of Frobenius can be used to 
find the coefficients $a_j$.
Plugging in the even-parity potential from 
Eq.~(\ref{eq:evenPotential}) a recurrence relation for the 
$a_j$ is
\begin{multline}
2 i \la^2 j \, a_{j} = 
 \la \Big[ \la \l j-1\r j 
- 12 i  \s  \l j -1 \r 
- 2 \la \l \la + 1 \r \Big] \, a_{j-1} \\
+ 2 \s \Big[  \la  \l 3 - \la \r \l j-2\r \l j-1 \r 
- \l \la^2 + 9 i \s  \r \l j-2 \r 
- 3 \la^2  \Big] \, a_{j-2} \\
+ 3 \s^2 \Big[ \l 3 - 4\la \r \l j-3\r \l j-2 \r 
- 4 \la \l j - 3 \r 
- 6 \la \Big] \,  a_{j-3} 
- 18 \s^3 \l j-3 \r^2 
\,  a_{j-4} 
\end{multline}
where $\s \equiv M \o_{mn}$.
For the odd-parity expansion, we plug in the potential in 
Eq.~(\ref{eq:oddPotential}).  The resulting recurrence relation is
\be
2 i j \, a_j  
 =  - 2 \s \Big[  \l j + 1 \r \l j - 3 \r \Big] \, a_{j - 2}  
- \Big[ \ell \l \ell + 1 \r - j \l j - 1 \r  \Big] \, a_{j-1} .
\ee
In order to use these recurrence relations,  the first few terms 
$a_0, \ a_1, \ldots $ are needed.  The recurrence relations 
actually provides them if one assumes that 
$a_j = 0$ for all negative $j$.

\bibliography{FrequencyMethods}

\end{document}